\documentclass[journal, 10pt, onecolumn]{IEEEtran}
\usepackage{graphicx}
\usepackage{float}
\usepackage{multicol} 
\usepackage{multirow}
\usepackage{times}
\usepackage{latexsym} 
\usepackage{array}    
\usepackage{subfigure}          % subfigure package
\usepackage{amsmath, amssymb, mathrsfs}
\usepackage{url}                % For email addresses, hypertext links, ...
\usepackage{ifpdf}              % For pdf and postscript files
\usepackage{pdfpages}           % For inserting pdf pages
\usepackage{algorithm}
\usepackage[noend]{algpseudocode}
\ifpdf
  \DeclareGraphicsExtensions{.pdf,.png,.jpg}
\else
  \DeclareGraphicsExtensions{.eps,.ps}
\fi

\renewcommand{\thesection}{\arabic{section}}
\renewcommand{\thesubsection}{\thesection.\arabic{subsection}}
\renewcommand{\thesubsubsection}{\thesubsection.\arabic{subsubsection}}

\usepackage{amssymb}
\usepackage{wrapfig}
\usepackage{comment}
\usepackage{balance}

\begin{document}
\title{\large{\bf A Network-of-Networks Model for Electrical Infrastructure Networks}}
\author{Mahantesh Halappanavar$^1$, Eduardo Cotilla-Sanchez$^2$, Emilie Hogan$^1$, 
   Daniel Duncan$^2$, \\ Zhenyu (Henry) Huang$^1$, and~Paul D.H.~Hines$^3$
\thanks{$^1$ Pacific Northwest National Laboratory.
        $^2$ Oregon State University.
        $^3$ University of Vermont. 
Email:\{hala@pnnl.gov, ecs@eecs.oregonstate.edu, emilie.hogan@pnnl.gov, duncanda@onid.orst.edu, 
        zhenyu.huang@pnnl.gov, paul.hines@uvm.edu\}. 
}}

\maketitle

%\tableofcontents

\begin{abstract}
Modeling power transmission networks is an important area of
research with applications such as vulnerability analysis,
study of cascading failures, and location of measurement devices. 
Graph-theoretic approaches have been widely used to solve these problems, 
but are subject to several limitations. 
One of the limitations is the ability to model a heterogeneous system 
in a consistent manner using the standard graph-theoretic formulation.
In this paper, we propose a {\em network-of-networks} approach for modeling 
power transmission networks in order to explicitly incorporate 
heterogeneity in the model. This model distinguishes between different 
components of the network that operate at different voltage ratings, and 
also captures the intra and inter-network connectivity patterns. 
By building the graph in this fashion we present a novel, and fundamentally 
different, perspective of power transmission networks. 
Consequently, this novel approach will have a significant impact on the 
graph-theoretic modeling of power grids that we believe will lead to a better 
understanding of transmission networks. 
\end{abstract}
\maketitle

\section{Introduction}
\label{sec:intro}
The topological structure of electric power grids can be modeled
as a graph. 
A graph $G=(V, E)$ is a pair, where the vertex set $V$ represents unique
entities and the edge set $E$ represents binary relationships on $V$.
The study of topological and electrical structure of 
power transmission networks is an important area of research 
with several applications such as vulnerability analysis~\cite{hines-chaos2010}, 
locational marginal pricing~\cite{DeMarco}, controlled islanding~\cite{Xu-2010}, 
and location of sensors~\cite{Anderson}.
In this paper, we focus on two interrelated problems: $(i)$ graph-based modeling 
and characterization of power transmission networks, and $(ii)$ random graph models 
for synthetic generation of transmission networks. 
These two problems have a significant impact on several aspects of our understanding 
and functioning of the power grid.

%%%%%%%%%%%%%%%%%%%%%%%%%%%%%%%%%%%%%%%%%%%%%%%%%%%%%%%%%%%%%%%%%%%%%%%%%%
%\begin{figure*}[h!]
\begin{wrapfigure}{r}{0.5\textwidth}
  \begin{center}
   \includegraphics[width=0.48\textwidth]{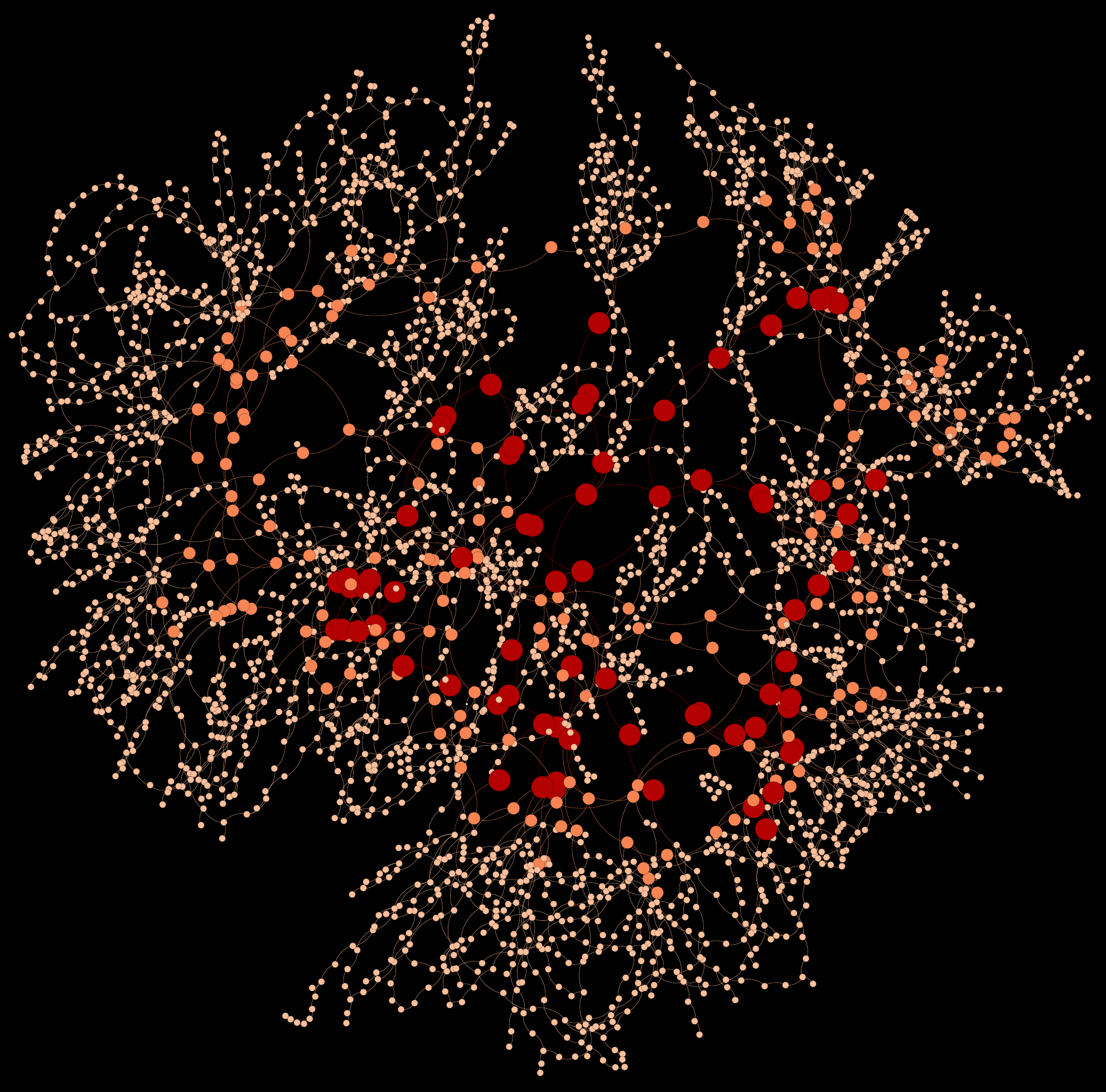}
   \end{center}
   \caption{Rendering of the Polish system with $3120$ vertices and $3684$ edges.
  The colors and sizes represent different voltage ratings. Larger the size, 
  higher the voltage. Details are provided in Section~\ref{sec:characterization}
  \vspace{0.3in}}
  \label{fig:PolishAll}
\end{wrapfigure}  
%\end{figure*}
%%%%%%%%%%%%%%%%%%%%%%%%%%%%%%%%%%%%%%%%%%%%%%%%%%%%%%%%%%%%%%%%%%%%%%%%%%
Graph-theoretic approaches have been studied extensively in the context of power 
grids (Section~\ref{sec:related}). 
However, there are several limitations that need to be addressed. 
For example, power grids have been characterized using graph-theoretic
metrics to show conflicting features~\cite{Cotilla-Sanchez-Compare}, and in the study 
of vulnerabilities to often misleading conclusions~\cite{hines-chaos2010}.  
Graph based models and algorithms are effective when vertices represent 
homogeneous entities in a consistent manner. However, {\em a transmission network is 
inherently heterogeneous}. It consists of entities such as generators, loads, 
substations, transformers, and transmission lines that operate at different 
voltage ratings. Further, power transformers are not always treated consistently. 
They are generally represented as edges, but not always.
Consequently, a purely graph-based model will misrepresent a transmission network when 
it is not constructed carefully. As an example, illustration of a model for the 
power grid in Poland is provided in Figure~\ref{fig:PolishAll}, where nodes operating 
at different voltages are shown in different colors and sizes -- larger the size, 
higher the nominal voltage of a node. Thus, edges connecting two different types of 
vertices are transformers. 
In this paper, we address the problem of accurate representation of power transmission 
networks as a graph, and discuss the impact of this representation for graph-theoretic characterization and modeling using random graph models.

\subsection{Contributions}
By accurately modeling transformers, and consequently decomposing a network into 
regions of different voltage ratings, we propose a novel 
network-of-networks model for power transmission networks. 
This simple idea has profound implications to the study of topological
and electrical structure of power grids. 

The contributions we make in this paper are:
$(i)$ a new decomposition method for power transmission networks 
using nominal voltage ratings of entities (Section~\ref{sec:characterization}); 
$(ii)$ empirical evidence of the hierarchical nature of transmission networks 
based on the analysis of real-world data representing the north American and 
European grids (Section~\ref{sec:characterization} and \ref{sec:interconnection});
$(iii)$ characterization of the interconnection structure between networks of 
different voltages (Section~\ref{sec:interconnection}); and
$(iv)$ presentation of random graph models -- two methods for modeling the network 
at a specific voltage level, and one method for modeling the interconnection structure 
of networks of different voltage (Section~\ref{sec:RandomGraphModel}).

\section{Network-of-Networks Model}
\label{sec:NonModel}
\begin{figure}[!htb]
\centering
\includegraphics[width=0.65\textwidth]{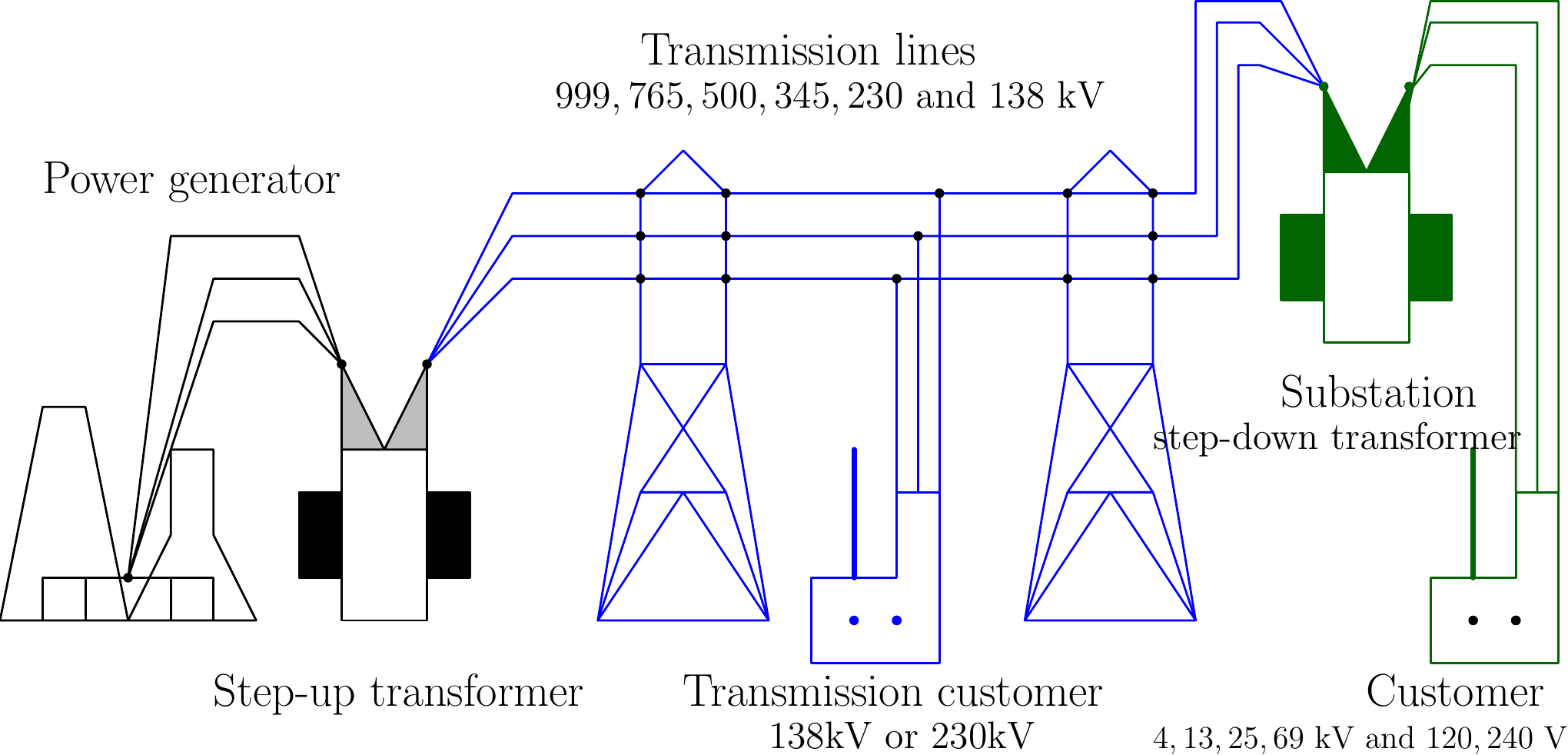}
\caption{\label{blockDiagram}
An illustration of a power system. Power generation, shown in black, is 
connected to distribution, shown in green, via a transmission system, 
shown in blue. The system operates at different voltage ratings as shown 
in the figure.}
\end{figure}

An electric power transmission network consists of power generators,
loads, transformers, substations, and branches connecting different
components.
While branches and transformers are modeled as edges of a graph, all
the other components (buses) are modeled as vertices. 
While this model is fairly consistent, the inclusion of transformers
as edges makes the graph representation inconsistent. 
An illustration of a symbolic  power system is shown in Figure~\ref{blockDiagram}. 
Generation is shown in black, transmission is shown in blue, and 
distribution in green. 
%We focus on the structure of transmission networks in this paper.
%%%%%%%%%%%%%%%%%%%%%%%%%%%%%%%%%%%%%%%%%%%%%%%%%%%%%%%%%%%%%%%%%%%%%%%
%%%%%%%%%%%%%%%%%%%%%%%%%%%%%%%%%%%%%%%%%%%%%%%%%%%%%%%%%%%%%%%%%%%%%%% 
A power transformer is used to step up or step down voltage between
its two end-points, and therefore, connects two different regions of
voltage magnitudes. 
Power transmitted at a certain voltage remains at that level
unless stepped up or stepped down by another transformer connecting the
network to a region of different voltage rating. 
Therefore, the component of the network transmitting power at a certain
voltage rating can be treated as one network where all the vertices
consistently represent entities operating at the same voltage rating, and 
edges represent transmission lines. 
The overall transmission network can thus be decomposed into a
{\em network-of-networks} (defined further below), where each network 
operates at a certain voltage rating, and a pair of networks are connected 
via one or more transformers regulating the voltage levels. 
When expressed in this manner, a different graph structure emerges, which
is the focus of our study. 

\subsection{Preliminaries}
\label{ssec:preliminaries}
A graph $G=(V, E)$ is a pair, where the vertex set $V$ represents unique
entities and the edge set $E$ represents binary relations on $V$. 
We define a network-of-networks model for power transmission as an
undirected heterogenous graph $G=(V, E, X, Y, \phi, \varphi)$ 
with a weight function $w: E \to \mathbb{R}^{+}$.
The vertex set $V$ represents unique entities such as substations, generators 
and loads; the edge set $E$ represents binary relationships on $V$ such 
that an edge $e \in E$ represents either a power transmission line or a 
voltage transformer; a set $X$ of values associated with $V$ representing 
the nominal voltage rating at a particular vertex as defined in the mapping 
function $\phi : V \to X$ where each vertex operates at a given voltage 
level $\phi (v)\in X$, and $X \in \mathbb{R}^{+}$ is a positive real number.
A set $Y$ of values associated with $E$ represent the type of an
edge -- a transmission line or a transformer -- as defined in the
mapping function $\varphi : E \to Y$ where each edge is of a given
type $\varphi (e) \in Y$, and $Y \in \mathbb{Z}$ is an integer. 

Traditionally, transmission networks are represented as graphs, 
making no distinction between entities operating at different voltage 
ratings and edges of different types.
In our representation, the vertex and edge types ($X, Y$), and
the mappings ($\phi, \varphi$), allow us to make clear distinctions
between different vertex and edge types. Given the voltage levels for
each substation ($X, \phi$), we can automatically detect the voltage
transformers, $E_T \subseteq E$, where $e_t(u, v) \in E_T$ such that
$\phi(u) \neq \phi(v)$. By pruning the transformer-edges from $G$, the
remaining graph represents a collection of subgraphs where each
subgraph has vertices operating at the same voltage level and edges
that are transmission lines. The structure of this collection of
subgraphs is of great interest in studying the transmission network and is
the focus of this paper.

{\bf Graph-theoretic Measures:} 
We use the following graph-theoretic measures in this paper. 
A {\em path} in a graph is a finite sequence of edges such that every edge 
in a path is traversed only once and any two consecutive edges in a path
are adjacent to each other (share a vertex in common). 
The {\em average shortest-path length} is defined as the average of 
shortest paths between all possible pairs of vertices in a graph.
The longest shortest-path is called the {\em diameter} of a graph.
The {\em local clustering coefficient} of a vertex $v$ is the ratio
between the actual number of edges among the neighbors of $v$ to the 
total possible number of edges among these neighbors. 
The {\em average clustering coefficient} of a graph is the sum of all local 
clustering coefficients of its vertices divided by the number of vertices.

\section{Characterization of Real-world Networks}
\label{sec:characterization}
We now present details of the network-of-networks model by 
decomposing different datasets using the method described in 
Algorithm~\ref{alg:decompose}.

%%%%%%%%%%%%%%%%%%%%%%%%%%%%%%%%%%%%%%%%%%%%%%%%%%%%%%%%%%%%%%%%%%%%%%%%%%%%%%%%
\begin{algorithm}[!htb]
  \caption{Decompose a graph based on voltage ratings. 
  {\bf Input:} A graph $G=(V, E, X, \phi)$.
  {\bf Output:} A graph $G'=\{G_i, G_j, ...\}$, which is a collection of disconnected 
  components of $G$.}
  \label{alg:decompose}
  \begin{algorithmic}[1]
  \Procedure{Decompose-Graph}{$G=(V, E, X, \phi), G'$}
   \State $Q \gets V$
   \State $G' \gets \emptyset$ \Comment{A collection of graphs}
   \While{$Q \neq \emptyset$}
     \State $v \gets Q \setminus \{v\}$ in some order
     \State $Q_t \gets \{v\}$
     \State $G_v \gets \{v\}$ \Comment{Represents a graph of same voltage level} 
     \While{$Q_t \neq \emptyset$}
    		\State $x \gets Q_t \setminus \{x\}$ in some order
    		\State $Q \gets Q \setminus \{x\}$
    		\For{{\bf each} $y \in $adj$(x)$} \Comment{Neighbors of $x$}
    	   	\If{$X[x] = X[y]$} \Comment{Same voltage rating}
    	      \State $Q_t \gets Q_t \cup \{y\}$
    	      \State $V(G_v) \gets V(G_v) \cup \{y\}$ \Comment{A new vertex}
    	      \State $E(G_v) \gets E(G_v) \cup \{(x, y)\}$ \Comment{A new edge}
    	   	\EndIf
   	 	\EndFor
    \EndWhile
    \State $G' \gets G_v$ 
  \EndWhile
  \EndProcedure
  \end{algorithmic}
\end{algorithm}
%%%%%%%%%%%%%%%%%%%%%%%%%%%%%%%%%%%%%%%%%%%%%%%%%%%%%%%%%%%%%%%%%%%%%%%%%%%%%%%%
 
Let $G=(V, E, X, \phi)$ be the input graph, and $G'$ be the 
decomposed graph that is computed as the output of 
Algorithm~\ref{alg:decompose}.
The algorithm starts by adding the vertices in $G$ to a queue $Q$ 
(Line $2$ in Algorithm~\ref{alg:decompose}) in an arbitrary order.
The {\bf while} loop on Line $4$ iterates until all the vertices in 
the queue have been processed.
Let us consider the process for an arbitrary vertex $v$ chosen from 
$Q$ (Line $5$).
Vertex $v$, which is added to a new queue $Q_t$, acts as the source of a 
new network (a connected component of $G$) consisting all the vertices 
connected to $v$ through edges of the same voltage.
The search for the network corresponding to $v$ is managed by adding and 
removing vertices from $Q_t$, and persists until $Q_t$ becomes empty 
(the {\bf while} loop on Line $8$). 
For each vertex $x$ in $Q_t$, we process all the neighbors of $x$ 
(Line $11$) to check if they are of the same voltage rating. 
The neighbors are added (Lines $14$ and $15$) to the network corresponding 
to vertex $v$ if true (same voltage). 
Once all the connected vertices are added to $G_v$, the network is added
to $G'$ (Line $16$).
The algorithm resumes from a new vertex that is chosen arbitrarily
until all the vertices in $Q$ have been processed and added to corresponding 
networks in $G'$. 

%%%%%%%%%%%%%%%%%%%%%%%%%%%%%%%%%%%%%%%%%%%%%%
\begin{table*}[!htb]
\centering
\caption{\label{dataset}Datasets used in this study. ASPL stands for Average
  Shortest Path Length.} 
\begin{tabular}{|l|r|r|r|r|} \hline
Dataset & (Vertices, Edges) & Clust. Coeff. & ASPL & Diameter \\ 
\hline\hline
Polish  & $(2383, 8155)$ & $0.012$ & $12.58$ & $30$ \\ \hline
Western Interconnect (WI) & $(15090, 18153)$ & $0.027$ & $24.33$ & $66$ \\ \hline
Eastern Interconnect (EI)& $(49597, 62985)$ & $0.071$ & $35.80$ & $96$ \\ \hline
Texas   Interconnect (TI)& $(4756, 5848)$ & $0.020$ & $17.06$ & $40$ \\ \hline
\end{tabular}
\end{table*}
%%%%%%%%%%%%%%%%%%%%%%%%%%%%%%%%%%%%%%%%%%%%%%

In Table~\ref{dataset} we list the real-world datasets that are used in this 
study. The table also provides details of the network (without decomposition) 
in graph-theoretic metrics described in Section~\ref{ssec:preliminaries}.
We study the three major components of the North American power grid -- 
the Eastern, Western and Texas Interconnects, and the Polish system. 
The Eastern Interconnect data is from a North American Electric Reliability 
Corporation (NERC) planning model for 2012. 
Data for Western and Texas Interconnects are from the Federal Energy Regulatory 
Commission (FERC) Form 715 filings in 2005.
The data for North American power grid is obtained through the U.S. Critical
Energy Infrastructure Information (CEII) request process.
Data for the Polish system is included with MATPOWER, an open source power 
system simulation package \cite{Zimmerman:2011}.
For all the datasets we perform several steps to clean the data for 
consistency. For example, we remove all the isolated vertices and edges -- vertices
with no edges incident on them, and edges disconnected from the graph.

For each input in the dataset, we provide details of the decomposed networks including information about the number of edges (power transformers) between networks of different voltage levels. Detailed graph measures are provided for the largest component in each level. In order to highlight the topological structure we also provide visual renderings of the networks at each level for all the inputs. With these details, we aim to support of observations that the topological structures are not only similar across different voltage levels, but also across different inputs.

\subsection{The Polish Network}
\label{ssec:Polish}
In the network-of-networks model for the Polish system, there are three
voltage levels: $400, 200$ and $100$ kV. The number of components (a network at a 
specific voltage level) are one, one and five respectively for $400, 200$ and $100$ kV.
The size ($|V|,|E|$) of the largest components are
$(50, 58),(135, 174)$ and $(2193, 2485)$ respectively in the same order of voltage levels.
For the largest component in each level, the average clustering coefficients are $0.141$, $0.032$ and $0.008$;
the average path lengths are $6.484, 7.899$ and $37.572$; and 
the diameters are $17, 20$ and $92$. 

\begin{figure*}[h!]
  \centering
  \subfigure[400 kV]{\includegraphics[width=0.30\textwidth]
    {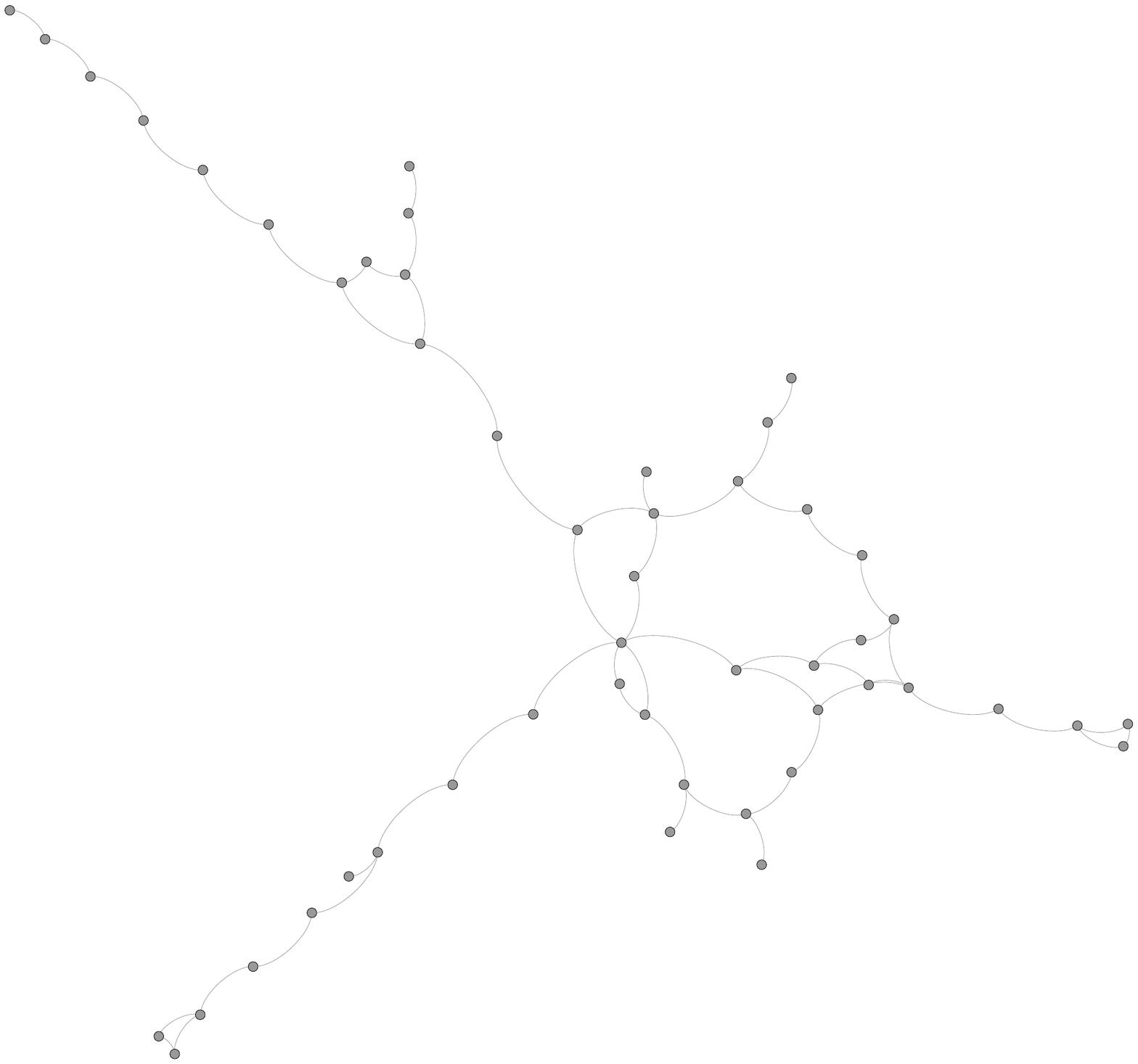}\label{f:Polish-400}}
  \subfigure[220 kV]{\includegraphics[width=0.30\textwidth]
    {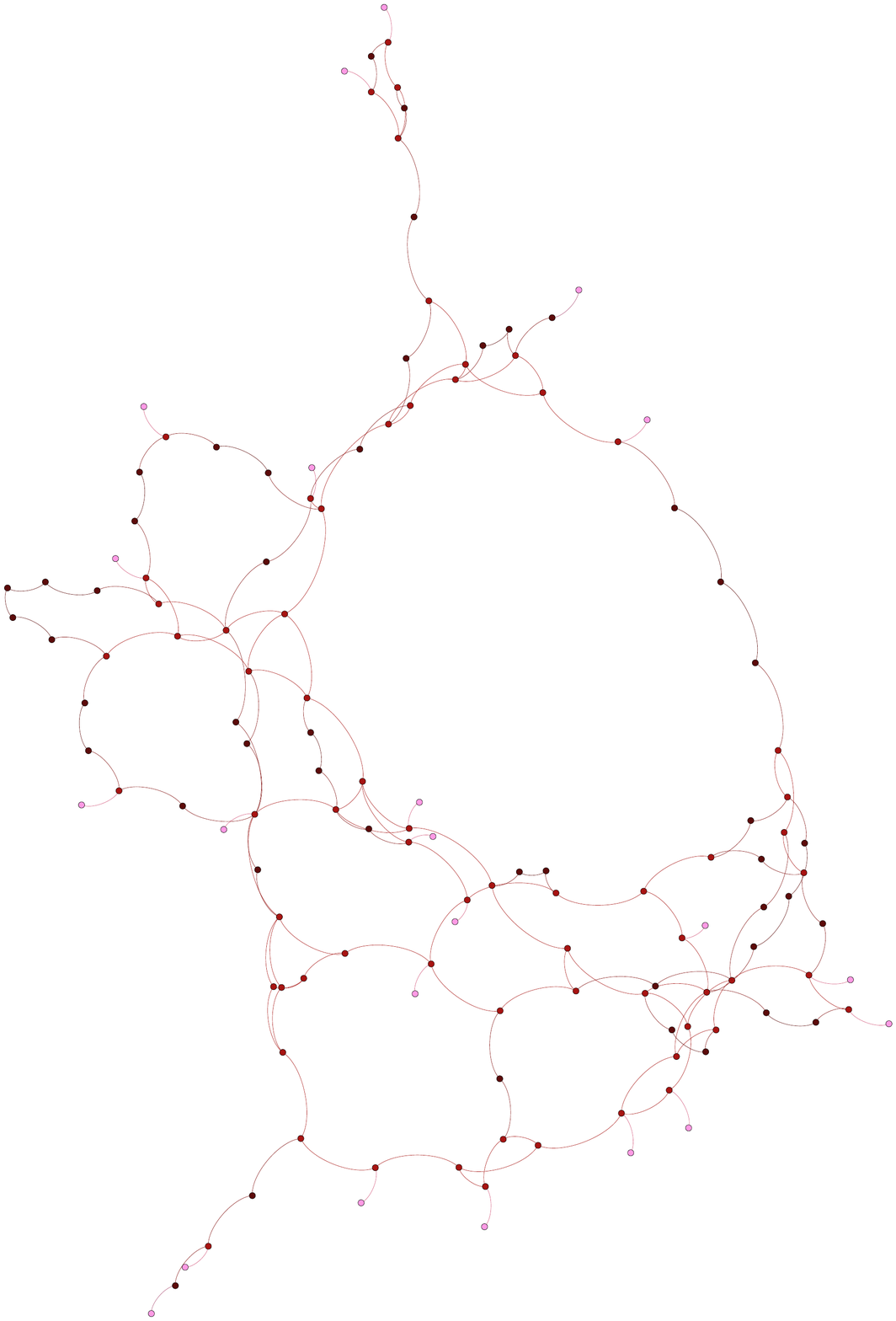}\label{f:Polish-220}}
  \subfigure[110 kV]{\includegraphics[width=0.30\textwidth]
    {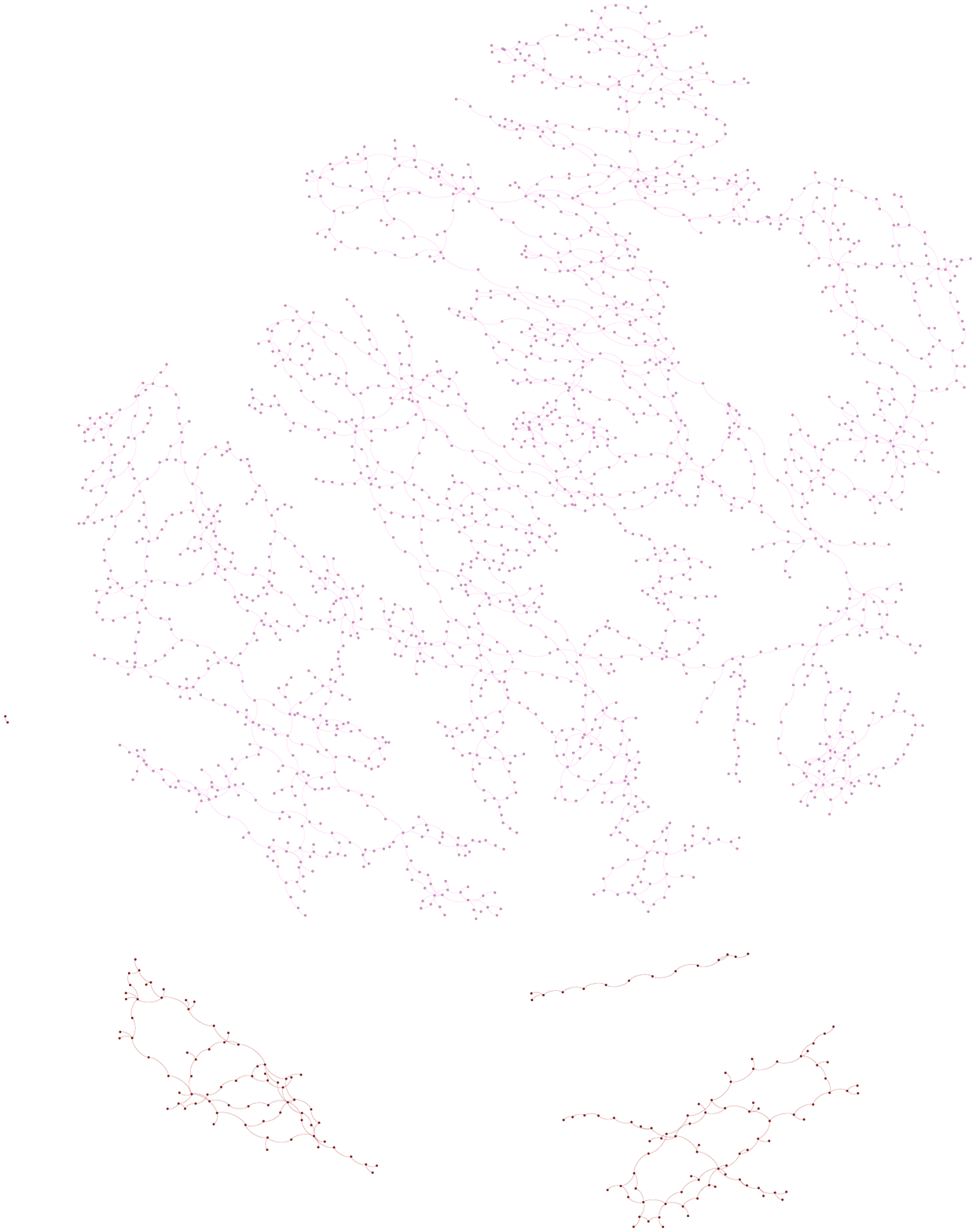}\label{f:Polish-110}}
  \caption{Rendering of the Polish network at different nominal voltage levels.}
  \label{fig:Polish}
\end{figure*}

Visualization of networks at different voltage levels of the Polish system is provided in Figure~\ref{fig:Polish}. 
We can visually observe from these renderings that the general topological structure consisting of chains and loops is consistent across different voltage levels. From an engineering perspective, loops provide alternative routes when one of the links in a chain fail, and chains help span large geographical areas.
%%%%%%%%%%%%%%%%%%%%%%%%%%%%%%%%%%%%%%%%%%%%%%%%%%%%%%%%%%%%%%%%%%%%%%%%%%
%%%%%%%%%%%%%%%%%%%%%%%%%%%%%%%%%%%%%%%%%%%%%%%%%%%%%%%%%%%%%%%%%%%%%%%%%%

In order to study the interconnection of different networks of the same voltage, we study the number of power transformers. The number of transformers (represented as edges) between voltage levels $400$ and $200$ kV is $15$, and that between $400$ and $100$ kV is $31$. There are $121$ transformers between voltage levels $200$ and $100$ kV. 

\subsection{Western Interconnect Network}
\label{ssec:WI}
%%%%%%%%%%%%%%%%%%%%%%%%%%%%%%%%%%%%%%%%%%%%%%%%%%%%%%%%%%%%%%%%%%%%%%%%%%
\begin{table*}[h!]
  \centering
  \caption{\label{wi_decompose}Details from decomposing the Western Interconnect dataset
  into four voltage levels. The values for average clustering coefficient, average shortest
  path length and diameter are for the largest component for a given voltage level.}
  \begin{tabular}{ | l | r | r | r | r | } \hline
    Voltage Levels & $500$ kV & $345$ kV & $230$ kV & $138$ kV\\\hline \hline
    Size ($|V|,|E|$) & $(384, 439)$ & $(305, 341)$ & $(1766, 2132)$ & $(5815, 6371)$\\\hline
    Num components & 4 & 12 & 20 & 104 \\\hline
    $|V|$ of largest component & 360 & 135 & 400 & 550 \\\hline
    Avg Clustering Coefficient & 0.02 & 0.047 & 0.055 & 0.022 \\\hline
    Avg Shortest Path Length & 22.397 & 9.571 & 20.693 & 39.822 \\\hline
    Diameter & 64 & 37 & 72 & 135 \\\hline
  \end{tabular}
\end{table*}
%%%%%%%%%%%%%%%%%%%%%%%%%%%%%%%%%%%%%%%%%%%%%%%%%%%%%%%%%%%%%%%%%%%%%%%%%%
The Western Interconnect has four voltage levels: $500, 345, 230$ and $138$ kV. 
Key metrics from different levels are summarized in Table~\ref{wi_decompose}.
A large part of the network operates at $138$ kV with $5815$ vertices and 
$6371$ edges. There are $104$ connected components (networks) at the $138$ kV, where the
largest component has $550$ vertices.  
The second largest portion of the network operates at $230$ kV consisting of $20$ components, with the largest component having an order of $1766$ vertices. 
The third largest portion operates at $500$ kV with $384$ vertices organized in four components. The remaining part of the network operates at $345$ kV with $305$ vertices divided over $12$ components. 
Different graph measures for the four voltage levels are proportional to the sizes of the corresponding graphs irrespective of the voltage ratings.
Accordingly, we observe diameters of $64, 37, 72$ and $135$ for the largest components in different voltage levels presented in descending order. While the average clustering coefficients vary, the average shortest path lengths are proportional to the size. 

\begin{figure*}[h!]
  \centering
  \subfigure[WI - 500 kV]{\includegraphics[width=0.48\textwidth]
    {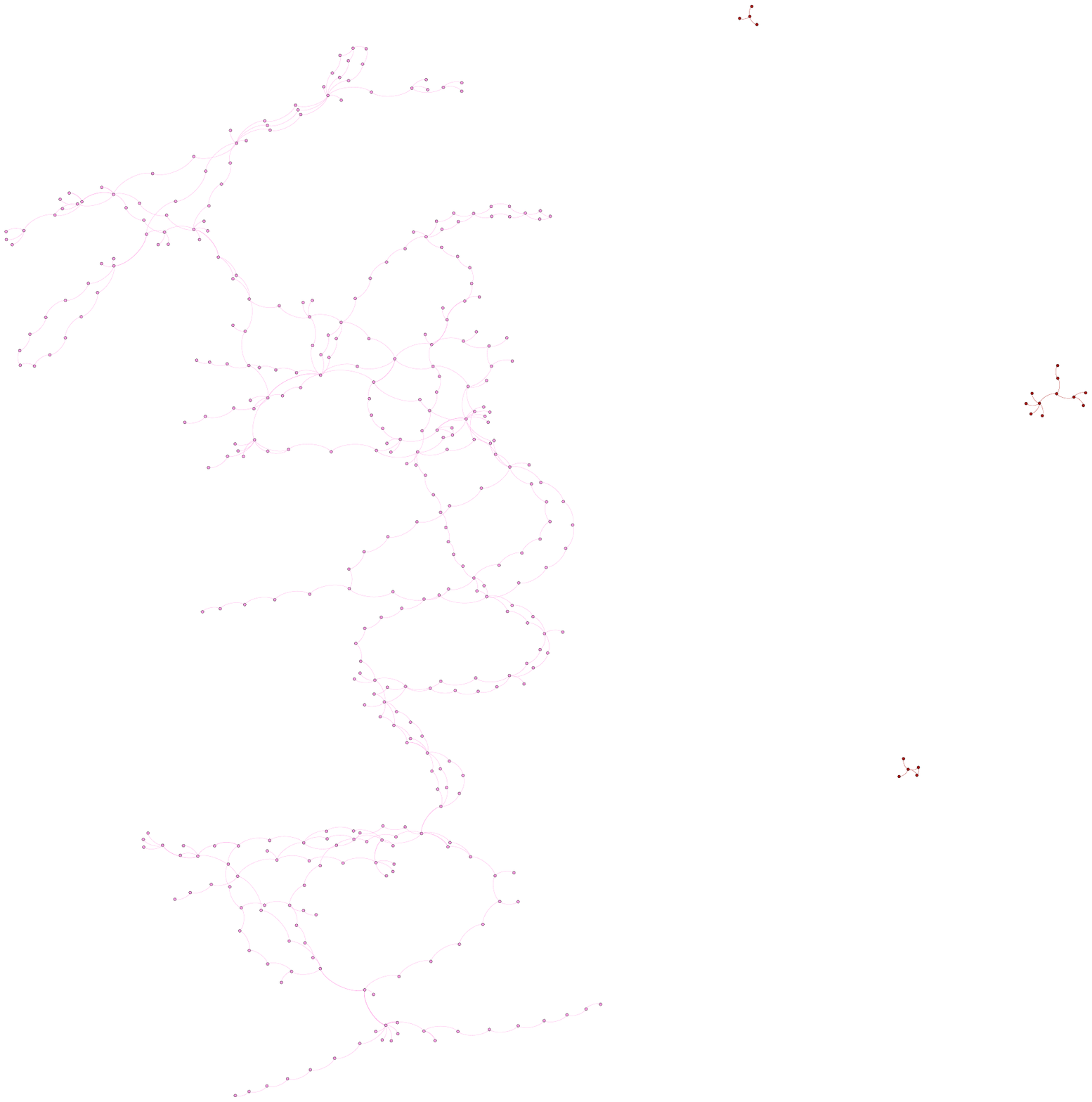}\label{f:WI-500}}
  \subfigure[WI - 345 kV]{\includegraphics[width=0.48\textwidth]
    {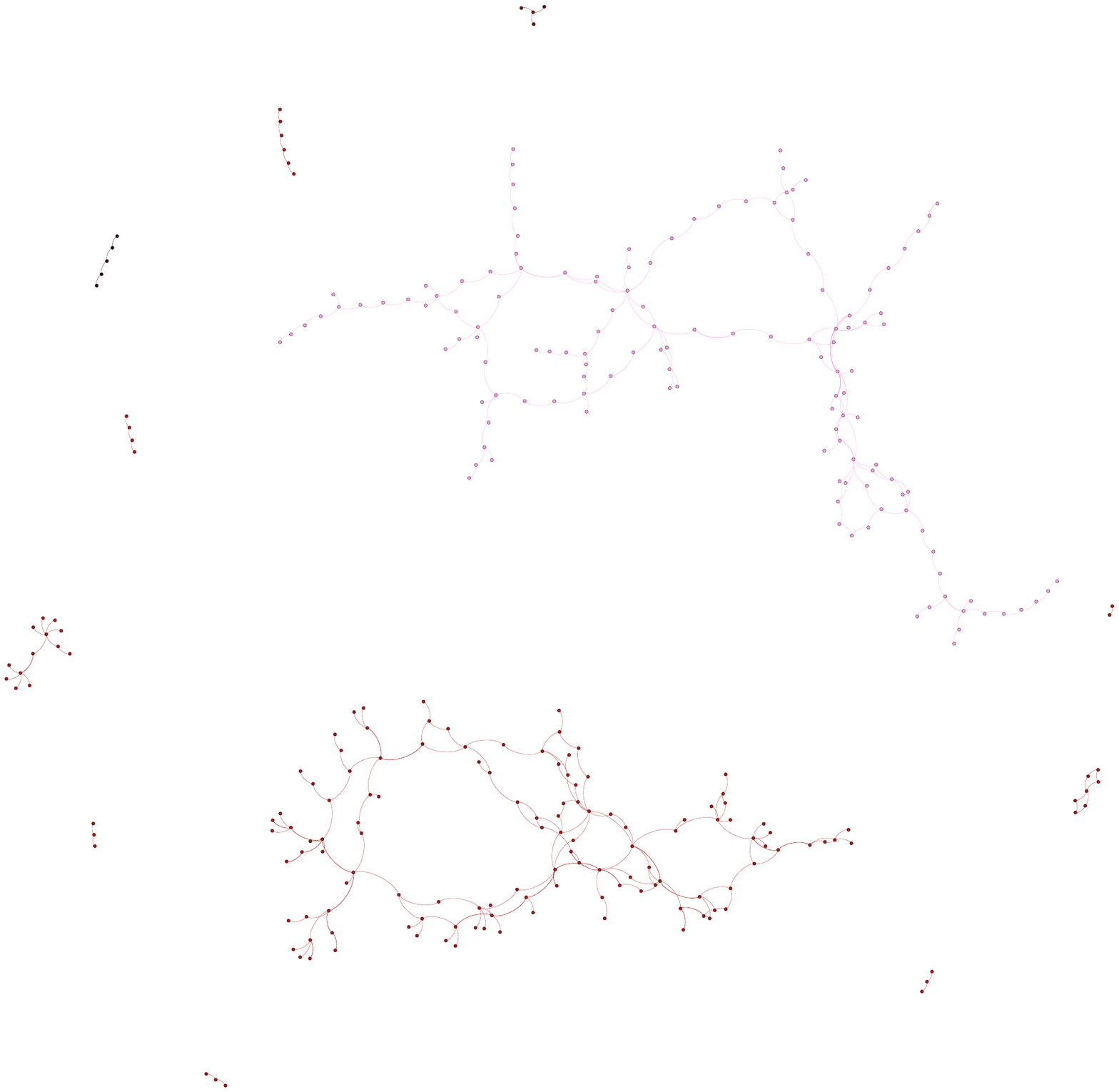}\label{f:WI-345}}
  \caption{Rendering of the Western Interconnect networks at $500$ and $345$ kV.}
  \label{fig:WI1}
\end{figure*}

\begin{figure*}[h!]
  \centering
  \subfigure[WI - 230 kV]{\includegraphics[width=0.48\textwidth]
    {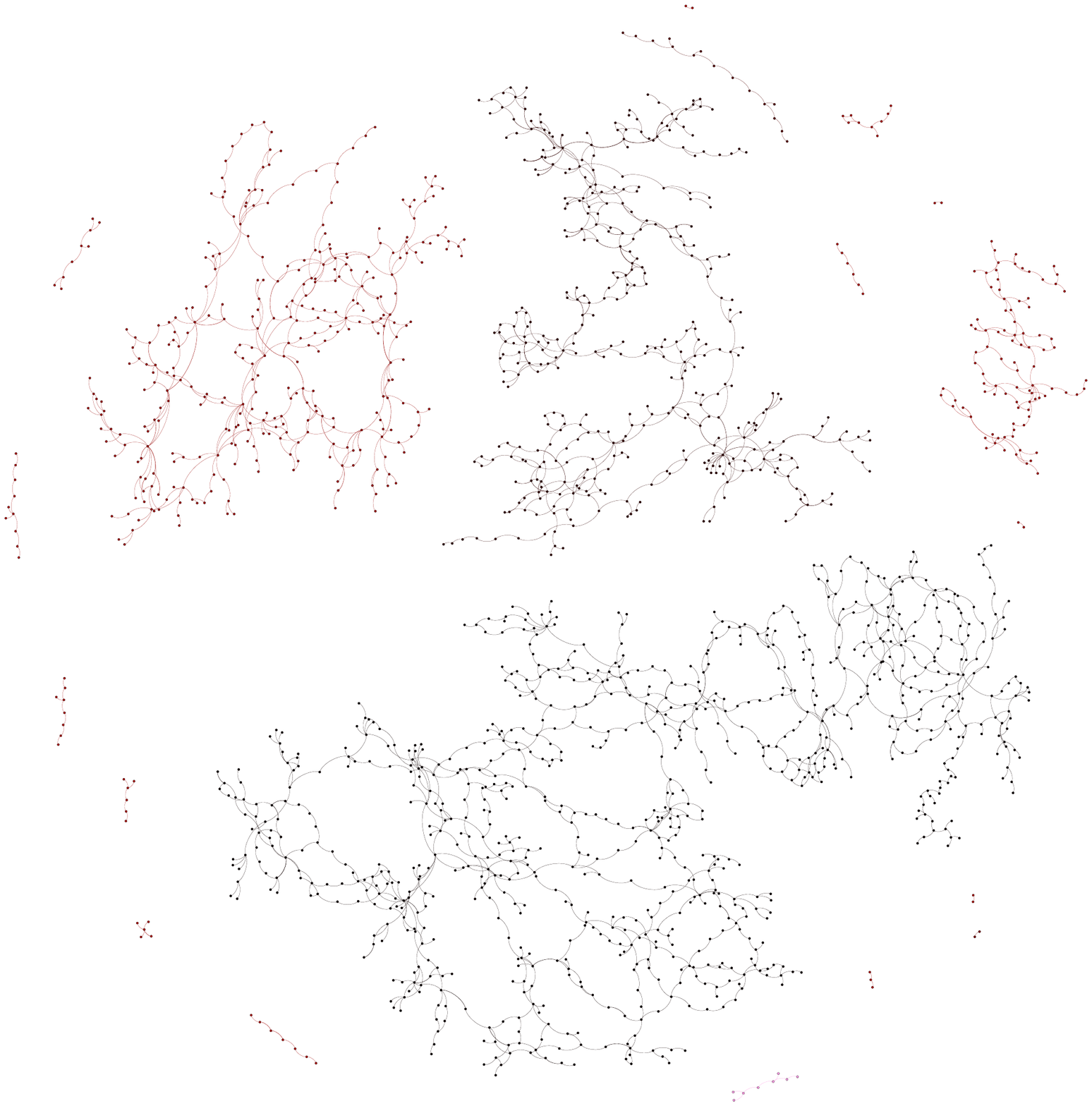}\label{f:WI-230}}
  \subfigure[WI - 138 kV]{\includegraphics[width=0.48\textwidth]
    {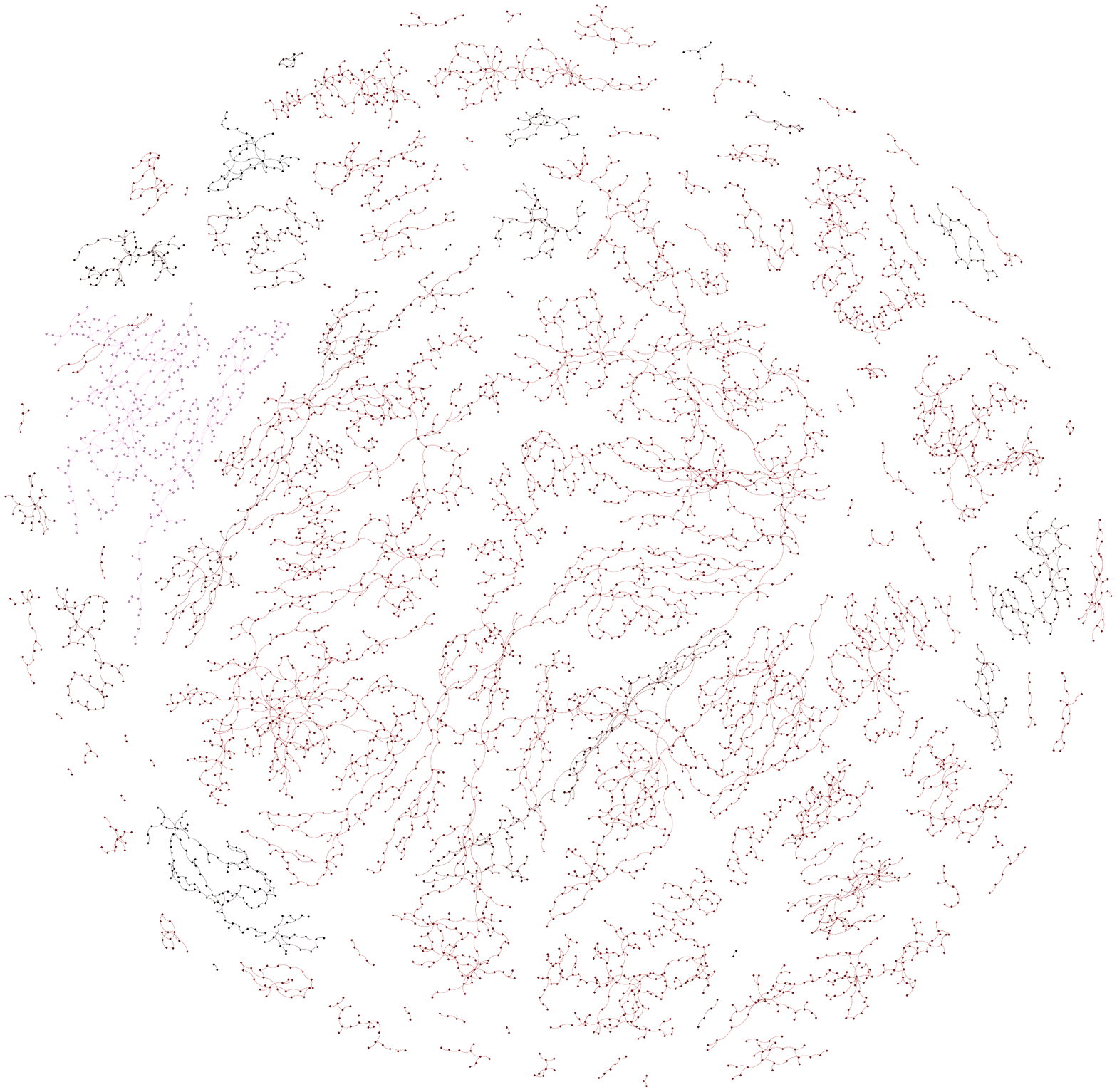}\label{f:WI-138}}
  \caption{Rendering of the Western Interconnect networks at $230$ and $138$ kV.}
  \label{fig:WI2}
\end{figure*}

Visualization of networks at voltage levels $500$ and $345$ kV is provided in Figure~\ref{fig:WI1}, and for $230$ and $138$ kV in Figure~\ref{fig:WI2}. Similar to networks in the Polish system, we observe that the pattern of loops and links is consistent across different voltage levels. Further, these patterns are also similar to the patterns
observed in the Polish system.
%%%%%%%%%%%%%%%%%%%%%%%%%%%%%%%%%%%%%%%%%%%%%%%%%%%%%%%%%%%%%%%%%%%%%%%%%%
%%%%%%% WECC 345 and 500 V %%%%%%%%
%%%%%%%%%%%%%%%%%%%%%%%%%%%%%%%%%%%%%%%%%%%%%%%%%%%%%%%%%%%%%%%%%%%%%%%%%%%%%%%%% 
%%%%%% WECC 138 and 230 V %%%%%%%
%%%%%%%%%%%%%%%%%%%%%%%%%%%%%%%%%%%%%%%%%%%%%%%%%%%%%%%%%%%%%%%%%%%%%%%%%%

The number of transformers (represented as edges) between different voltage levels for the Western Interconnect system is summarized in Table~\ref{wi_edges}. 
We observe that the maximum number of connections ($549$) exist between levels
$138$ and $230$ kV; followed by $123$ edges between levels  $138$ and $345$ kV; and
$115$ edges between levels  $230$ and $500$ kV. There are $10$ transformers between 
$500$ and $345$ kV. 
The manner in which these transformers are connected provides a rich interaction network.
Details and visualization of this interconnection structure is presented in Section~\ref{sec:interconnection}.
%%%%%%%%%%%%%%%%%%%%%%%%%%%%%%%%%%%%%%%%%%%%%%%%%%%%%%%%%%%%%%%%%%%%%%%%%%
\begin{table}[h!]
  \centering
  \caption{\label{wi_edges}Transformers in the Western Interconnect system}
  \begin{tabular}{ | l | r | r | r | r |} \hline
        & $500$ kV & $345$ kV & $230$ kV & $138$ kV \\ \hline \hline
    $500$ kV & -- & 10 & 115 & 18    \\ \hline
    $345$ kV & 10 & -- & 54 & 123    \\ \hline
    $230$ kV & 115 & 54 &  --  & 549 \\ \hline
    $138$ kV & 18 & 123 &  549  & -- \\ \hline
  \end{tabular}
\end{table}
%%%%%%%%%%%%%%%%%%%%%%%%%%%%%%%%%%%%%%%%%%%%%%%%%%%%%%%%%%%%%%%%%%%%%%%%%%

\subsection{Eastern Interconnect Network}
\label{ssec:EI}
\begin{figure*}[h!]
  \centering
  \subfigure[EI - 345 kV]{\includegraphics[width=0.48\textwidth]
    {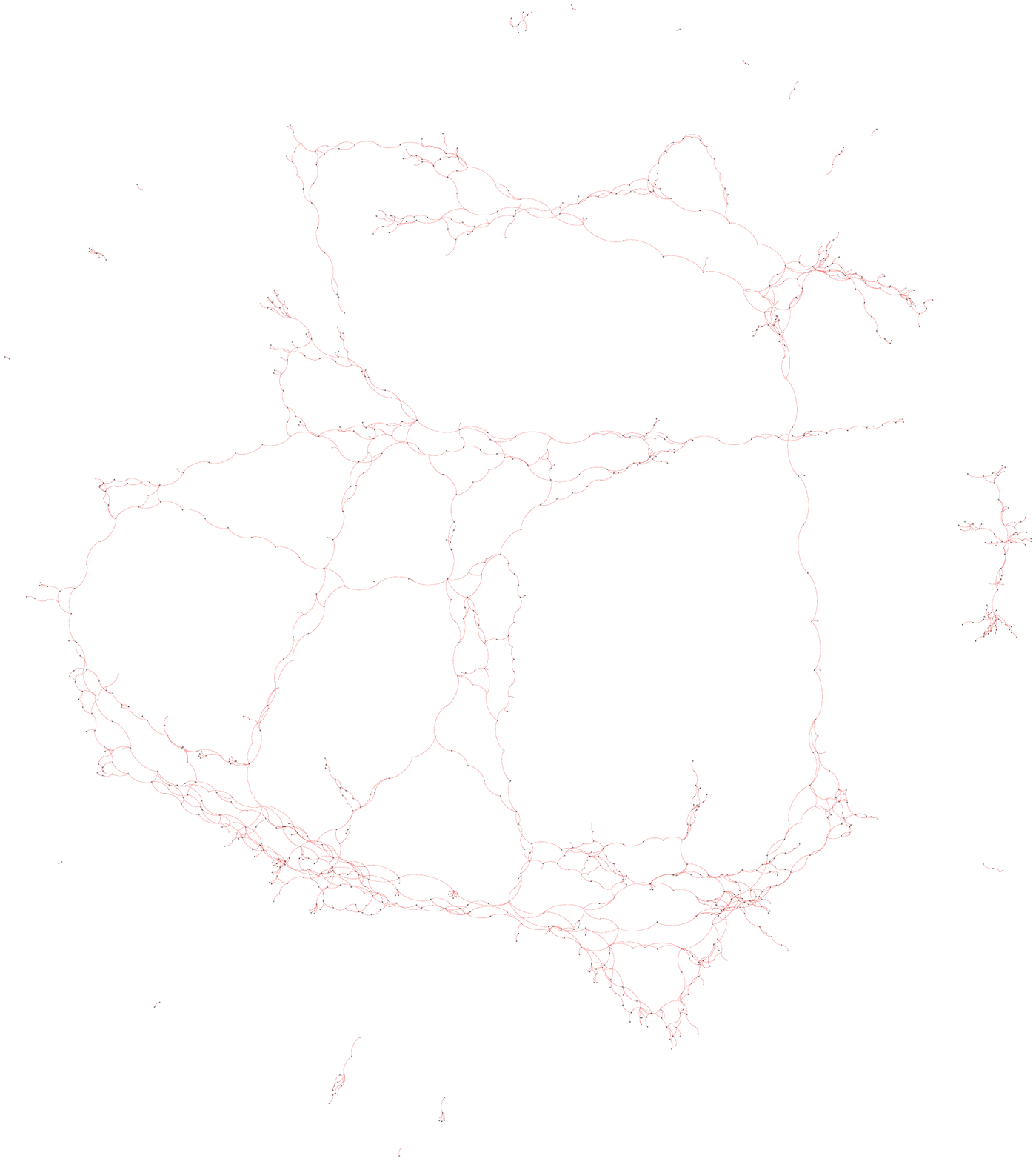}\label{f:EI-345}}
  \subfigure[EI - 230 kV]{\includegraphics[width=0.48\textwidth]
    {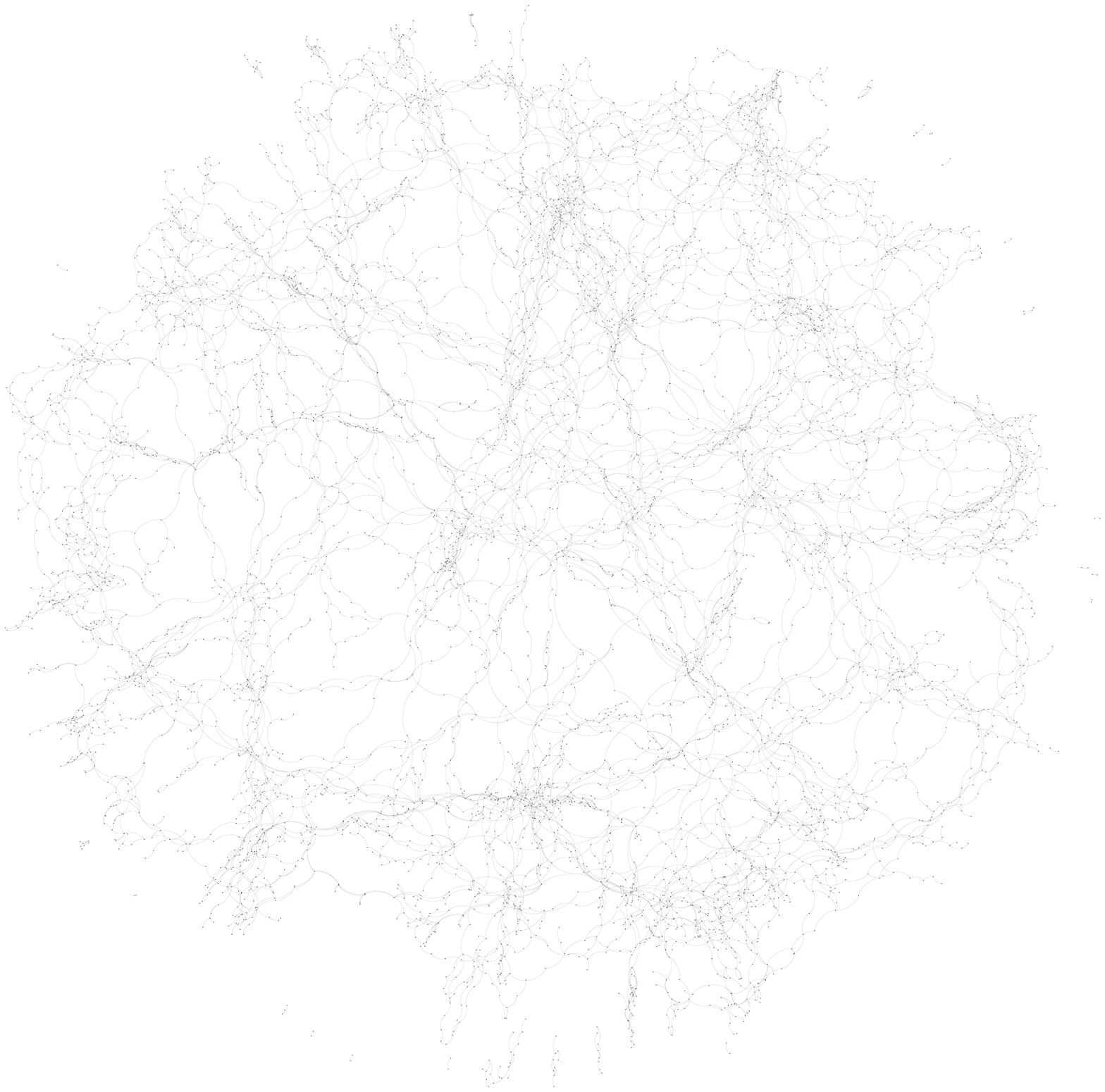}\label{f:EI-230}}
  \caption{Rendering of the Eastern Interconnect networks at $345$ and $230$ kV.}
  \label{fig:EI2}
\end{figure*}

The Eastern Interconnect is the largest instance in our dataset and has five voltage levels: $765, 500, 345, 230$ and $138$ kV. 
Key details of the system are summarized in Table~\ref{ei_decompose}.
The largest portion of the network operates at $138$ kV with $16082$ vertices and 
$19110$ edges. At this level, there are $121$ connected components with the
largest component comprising of $12997$ vertices. As a result, a large fraction of components at this level are small in size. A similar pattern also emerges at $230$ kV with $41$ components (Figure~\ref{f:EI-230}).

\begin{figure*}[h!]
  \centering
  \subfigure[EI - 765 kV]{\includegraphics[width=0.48\textwidth]
    {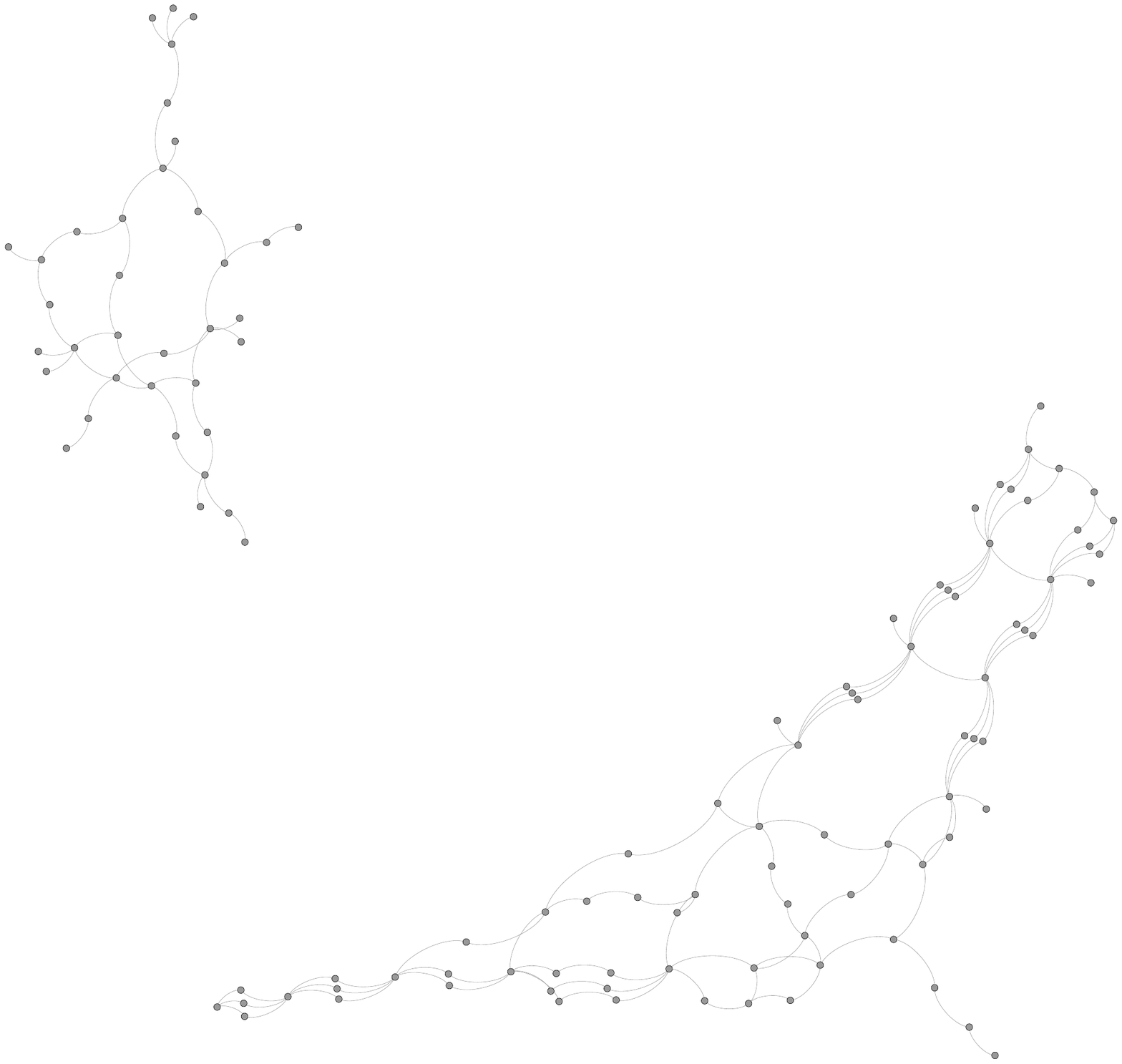}\label{f:EI-765}}
  \subfigure[EI - 500 kV]{\includegraphics[width=0.48\textwidth]
    {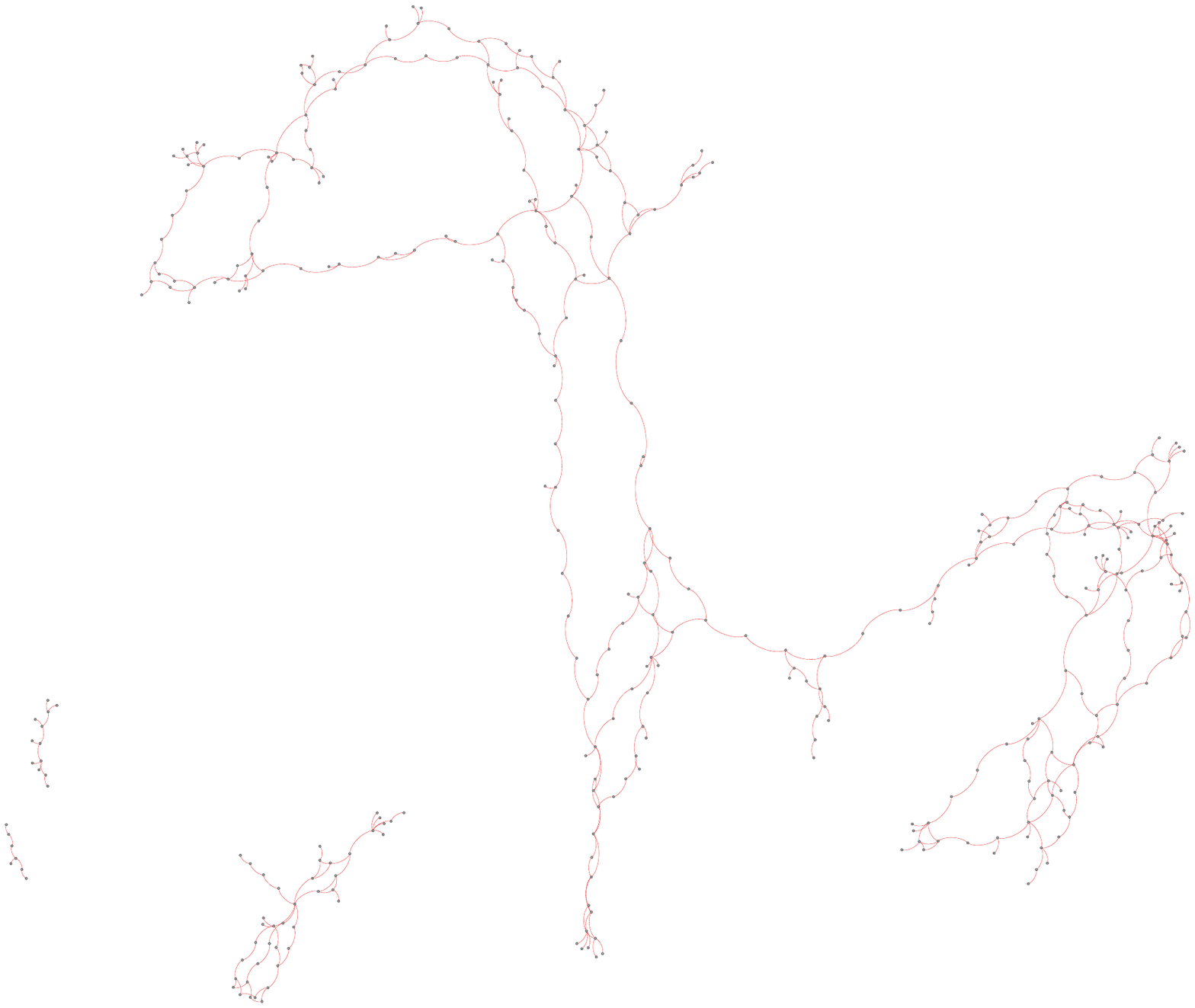}\label{f:EI-500}}
  \caption{Rendering of the Eastern Interconnect networks at $765$ and $500$ kV.}
  \label{fig:EI1}
\end{figure*}

Similar to the Western Interconnect, different graph measures for the five voltage levels are proportional to the sizes of the corresponding graphs irrespective of the voltage ratings.
We observe diameters of $386, 211, 78, 48$ and $16$ for the largest components in different voltage levels in descending order of kV. Again, the average clustering coefficients vary, but the average shortest path lengths are proportional to the size of the largest component. 
Rendering of the networks at voltage levels $765$ and $500$ kV is provided in Figure~\ref{fig:EI1}, and for $345$ and $230$ kV in Figure~\ref{fig:EI2}.
%%%%%%%%%%%%%%%%%%%%%%%%%%%%%%%%%%%%%%%%%%%%%%%%%%%%%%%%%%%%%%%%%%%%%%%%%%
\begin{table*}[h!]
  \centering
  \caption{\label{ei_decompose}Details from decomposing the Eastern Interconnect dataset
  into five voltage levels. The values for average clustering coefficient, average shortest
  path length and diameter are for the largest component for a given voltage level.}
  \begin{tabular}{ | l | r | r | r | r | r | } \hline
    Voltage Levels & $765$ kV & $500$ kV & $345$ kV & $230$ kV & $138$ kV\\ \hline \hline
    Size ($|V|,|E|$) &  $(115, 152)$  &   $(370, 441)$  & $(1224,
    1521)$ & $(6251, 7435)$ & $(16082, 19110) $\\ \hline
    $|V|$ of largest component  & $79$ & $310$ & $1065$ & $5688$ & $12997$\\ \hline
    Num components & 2  & 4 & 18 & 41 & 121 \\ \hline
    Avg Clustering Coefficient & 0.028   & 0.086 & 0.073 & 0.035 & 0.047 \\ \hline
    Avg Shortest Path Length & 5.93 & 18.68 &  24.098 & 73.453 & 134.84 \\ \hline
    Diameter & 16 & 48 & 78 & 211 & 386 \\ \hline
  \end{tabular}
\end{table*}
%%%%%%%%%%%%%%%%%%%%%%%%%%%%%%%%%%%%%%%%%%%%%%%%%%%%%%%%%%%%%%%%%%%%%%%%%%

%%%%%%%%%%%%%%%%%%%%%%%%%%%%%%%%%%%%%%%%%%%%%%%%%%%%%%%%%%%%%%%%%%%%%%%%%%
%%%%%%% EI 500 V and 765V %%%%%%%%
%%%%%%%%%%%%%%%%%%%%%%%%%%%%%%%%%%%%%%%%%%%%%%%%%%%%%%%%%%%%%%%%%%%%%%%%%%
%%%%%%% EI 230 V and 345V %%%%%%%%
%%%%%%%%%%%%%%%%%%%%%%%%%%%%%%%%%%%%%%%%%%%%%%%%%%%%%%%%%%%%%%%%%%%%%%%%%%

The number of transformers (represented as edges) between different voltage levels for the Eastern Interconnect system is summarized in Table~\ref{ei_edges}. 
Similar to the Western Interconnect, the maximum number of edges ($1093$) exist between the two lowest levels of voltage -- $138$ and $230$ kV. Visualization of the interconnection structure is discussed in Section~\ref{sec:interconnection}.
%%%%%%%%%%%%%%%%%%%%%%%%%%%%%%%%%%%%%%%%%%%%%%%%%%%%%%%%%%%%%%%%%%%%%%%%%%
\begin{table}[h!]
  \centering
  \caption{\label{ei_edges}Transformers in the Eastern Interconnect system}
  \begin{tabular}{ | l | r | r | r | r | r |} \hline
    & $765$ kV & $500$ kV & $345$ kV & $230$ kV & $138$ kV \\ \hline \hline
    $765$ kV  & -- & 5 & 38 & 13  & 18   \\ \hline
    $500$ kV  & 5 &  --  & 9 & 194 & 33   \\ \hline
    $345$ kV  & 38 & 9 & -- & 166 & 602  \\ \hline
    $230$ kV  & 13 & 194 & 166 & -- & 1093 \\ \hline
    $138$ kV  & 18 & 33 & 602 & 1093 & -- \\ \hline
  \end{tabular}
\end{table}
%%%%%%%%%%%%%%%%%%%%%%%%%%%%%%%%%%%%%%%%%%%%%%%%%%%%%%%%%%%%%%%%%%%%%%%%%%

\subsection{Texas Interconnect Network}
\label{ssec:TI}
\begin{figure*}[h!]
  \centering
  \subfigure[TI - 345 kV]{\includegraphics[width=0.30\textwidth]
    {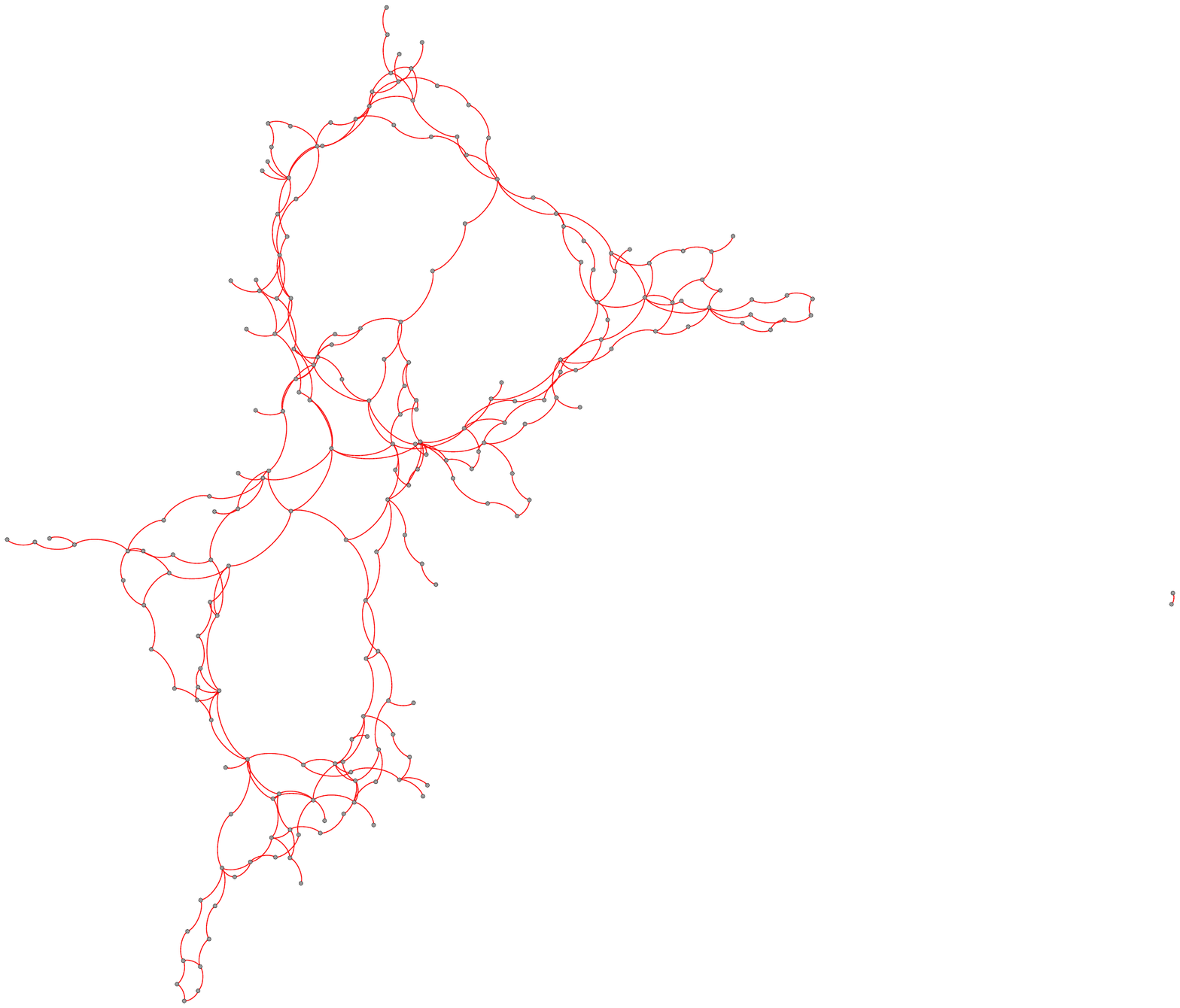}\label{f:TI-345}}
  \subfigure[TI - 138 KV]{\includegraphics[width=0.30\textwidth]
    {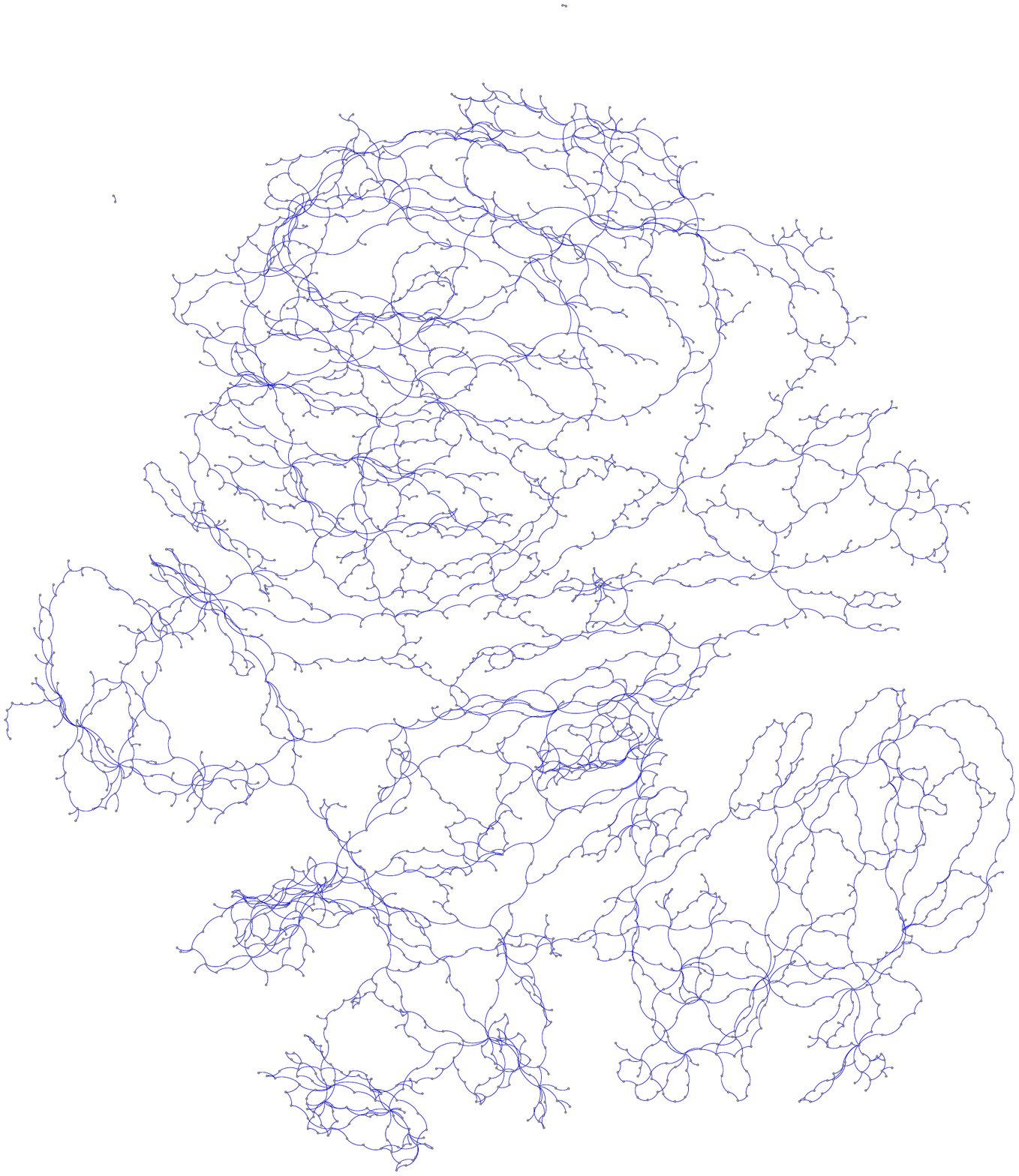}\label{f:TI-138}}
    \subfigure[TI - 69 KV]{\includegraphics[width=0.30\textwidth]
    {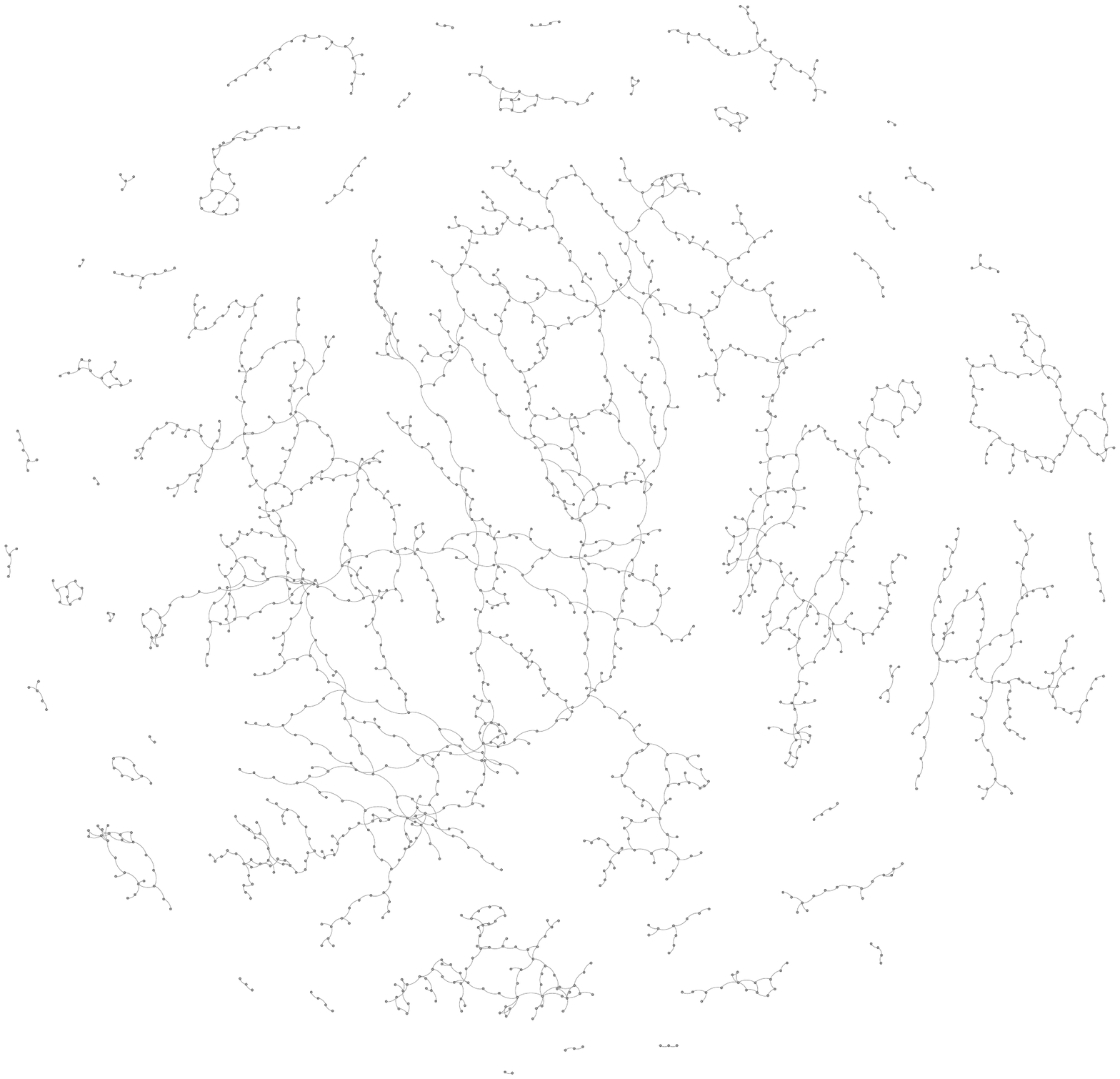}\label{f:TI-69}}
  \caption{Rendering of the Texas Interconnect network at $345, 138$ and $69$ kV.}
  \label{fig:TI}
\end{figure*}

The Texas Interconnect has three significant voltage levels: $345, 138$ and $69$ kV. 
Key details of the system are summarized in Table~\ref{ti_decompose}.
The largest portion of the network operates at $138$ kV with $2772$ vertices and 
$3274$ edges. At this level, there are $3$ connected components with the
largest component comprising of $2770$ vertices. The second largest portion of the network operates at $69$ kV, with $1749$ vertices and $1845$ edges. Unlike the network at $138$ kV, the network at $69$ kV has significantly larger number of connected components -- $44$ components. Consequently, the existence of a large number of small components are visible in Figure~\ref{f:TI-69}. Rendering of the networks at voltage levels $345, 138$ and $69$ kV is provided in Figure~\ref{fig:TI}.
%%%%%%%%%%%%%%%%%%%%%%%%%%%%%%%%%%%%%%%%%%%%%%%%%%%%%%%%%%%%%%%%%%%%%%%%%%
\begin{table*}[!htb]
  \centering
  \caption{\label{ti_decompose}Details from decomposing the Texas Interconnect dataset
  into three voltage levels. The values for average clustering coefficient, average shortest
  path length and diameter are for the largest component for a given voltage level.}
  \begin{tabular}{ | l | r | r | r | } \hline
    Voltage Levels & $345$ kV & $138$ kV & $69$ kV \\ \hline \hline
    Size $(|V|,|E|)$ &  $(210, 291)$ & $(2772, 3274)$ & $(1749, 1845)$\\ \hline
    Order of largest component  & $208$  & $2770$ & $1100$ \\ \hline
    Num components & 2  & 3 & 44  \\ \hline
    Avg Clustering Coefficient & 0.062  & 0.019   & 0.02 \\ \hline
    Avg Path Length & 8.32 & 32.466 & 47.22  \\ \hline
    Diameter & 22  & 84 & 134  \\ \hline
  \end{tabular}
\end{table*}
%%%%%%%%%%%%%%%%%%%%%%%%%%%%%%%%%%%%%%%%%%%%%%%%%%%%%%%%%%%%%%%%%%%%%%%%%%

%%%%%%%%%%%%%%%%%%%%%%%%%%%%%%%%%%%%%%%%%%%%%%%%%%%%%%%%%%%%%%%%%%%%%%%%%%
%%%%%%% TI 345, 138 and 69kV %%%%%%%%
%%%%%%%%%%%%%%%%%%%%%%%%%%%%%%%%%%%%%%%%%%%%%%%%%%%%%%%%%%%%%%%%%%%%%%%%%%

We conclude this section by highlighting the existence of a similar topological structure not only across different voltage levels, but also across different systems. The structure of a network at a given voltage level, decomposed using Algorithm~\ref{alg:decompose}, provides insight about the existence of large diameter (and average shortest path length) and low clustering coefficient in power transmission networks.

\section{Interconnection of networks}
\label{sec:interconnection}
Given a network-of-networks (decomposed) model of transmission network,
we compute the interconnection structure of the networks operating 
at different voltage levels using Algorithm~\ref{algo.interconnect}.
Let $G'=\{G_i, G_j, ...\}$ be the input (decomposed) graph as computed from 
Algorithm~\ref{alg:decompose}. 
Let $G''=(V, E, X, \phi)$ be the output graph representing the interconnection 
structure of the decomposed graph $G'$.
%%%%%%%%%%%%%%%%%%%%%%%%%%%%%%%%%%%%%%%%%%%%%%%%%%%%%%%%%%%%%%%%%%%%%%%%%%%%%%%%%%%%%
%%%%%%%%%%%%%%%%%%%%%%%%%%%%%%%%%%%%%%%%%%%%%%%%%%%%%%%%%%%%%%%%%%%%%%%%%%%%
\begin{algorithm}[!htb]
\caption{Interconnection structure}
\label{algo.interconnect}
\begin{algorithmic}[1]
\Procedure{Build-Interconnect-Graph}{$G'$, $G''$}
\For {{\bf each} $G_i \in G'$} \Comment{Add the vertices to $G''$}
    \State $V(G'') \gets V \cup \{v_i\}$ \Comment{Add a vertex $v_i$ representing component $G_i$}
    \State $\phi(G'') \gets \phi \cup \{v_i \to X(G_i)\}$ \Comment{Voltage rating of  component $G_i$}
\EndFor

\For {{\bf each} $G_i \in G'$} \Comment{Add the edges to $G''$}
	\For {{\bf each} $G_j \in G'$ such that $i \neq j$} 
         \State $E(G'') \gets E \cup \{(v_i, v_j)\}$ \Comment{Vertices $v_i$ and $v_j$ represent components $G_i$ and $G_j$ respectively}
    \EndFor
\EndFor

\EndProcedure
\end{algorithmic}
\end{algorithm}
%%%%%%%%%%%%%%%%%%%%%%%%%%%%%%%%%%%%%%%%%%%%%%%%%%%%%%%%%%%%%%%%%%%%%%%%%%%%%%%%%%%%%

Algorithm~\ref{algo.interconnect} proceeds by assigning a new vertex to each component
$G_i \in G'$. The voltage rating of $G_i$ is assigned to the representing 
vertex. An edge is added between two vertices if the corresponding components are 
interconnected via a power transformer. Consequently, note that in $G''$, each vertex represents a connected component of the same voltage rating, and each edge represents a 
transformer in the original graph $G$. Thus, the graph $G''$ represents how different regions of similar voltage are interconnected with each other. 

\begin{figure*}[h!]
  \centering
  \subfigure[Western Interconnect]{\includegraphics[width=0.48\textwidth]{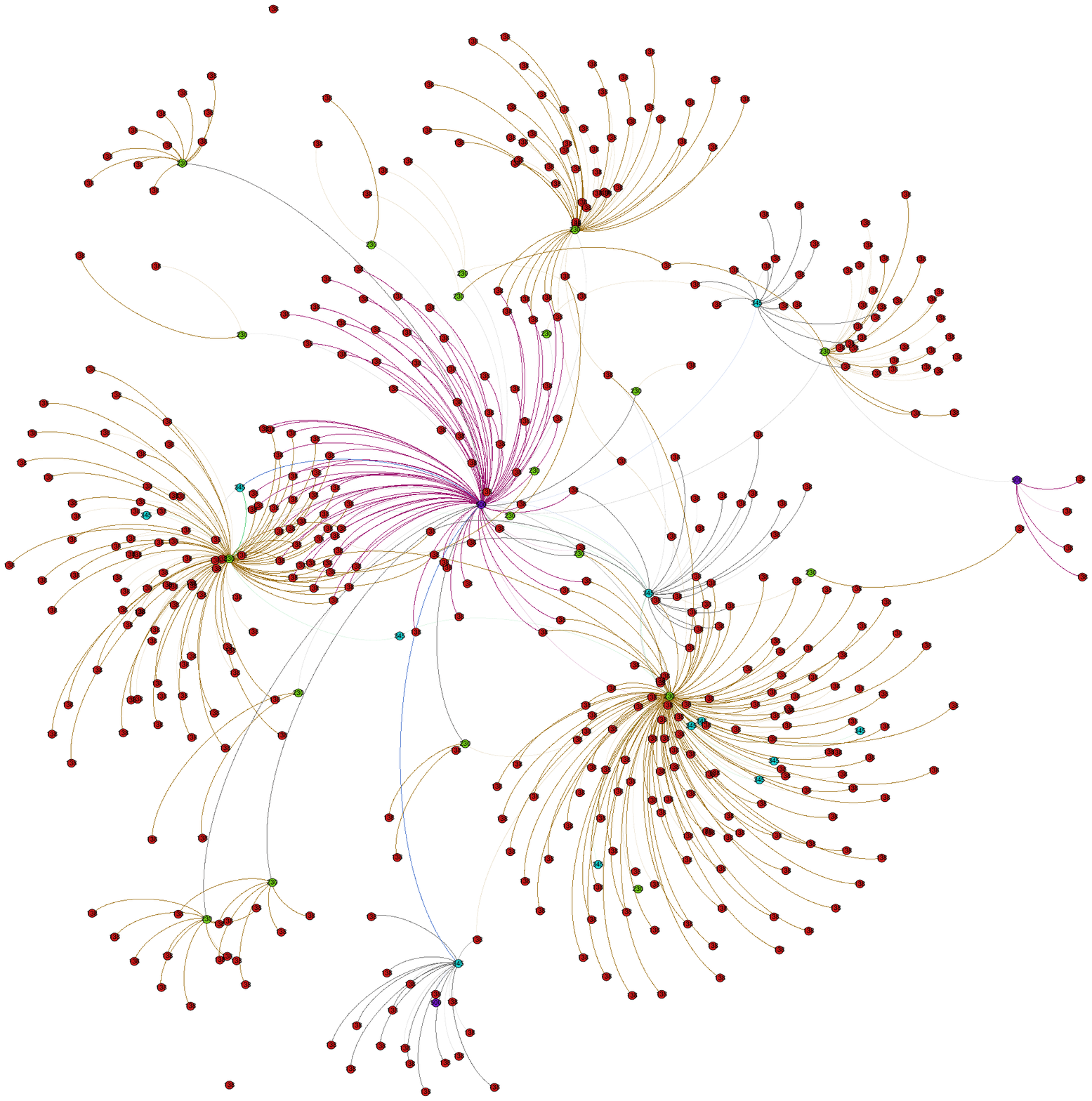}\label{f:WI-Interconnect}}
  \subfigure[Eastern Interconenct]{\includegraphics[width=0.48\textwidth]{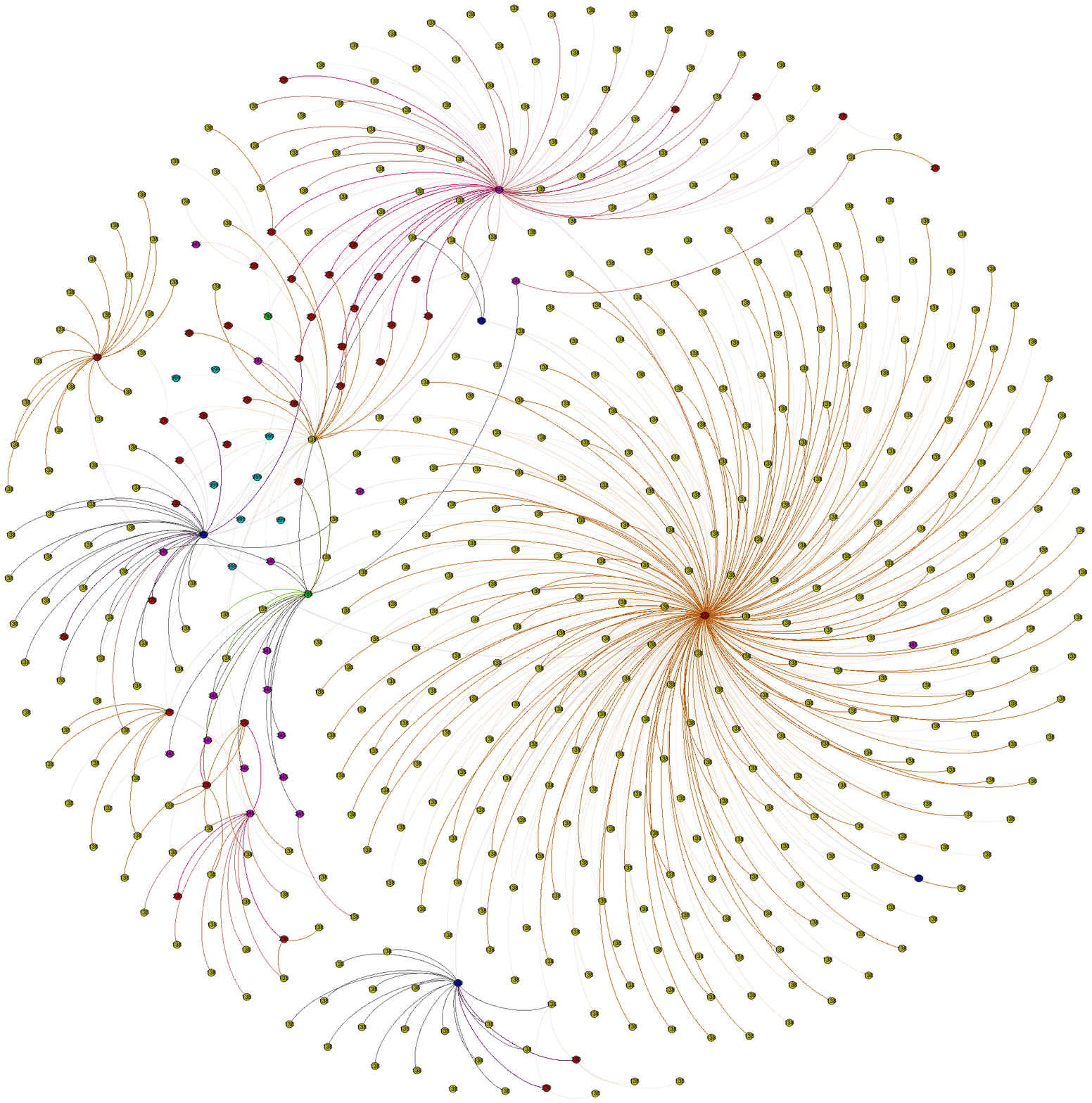}\label{f:EI-Interconnect}}
  \caption{Rendering of the interconnection network for the Western and Eastern interconnects. For the Western Interconnect, vertices in red color represent networks at $138$ kV, green at $230$ kv, blue at $345$ kV and purple at $500$ kV. 
  For the Eastern Interconnect, the color code is as follows: $138$ -- yellow green, $230$ -- red, $345$ -- purple, $500$ -- blue, and $765$ -- green.}
  \label{fig:interconnect}
\end{figure*}

%%%%%%%%%%%%%%%%%%%%%%%%%%%%%%%%%%%%%%%%%%%%%%
\begin{table*}[h!]
\centering
\caption{\label{interconnect}Interconnection networks. ASPL stands for Average
  Shortest Path Length.} 
\begin{tabular}{|l|r|r|r|r|}  \hline
Dataset & (Vertics, Edges) & Clust. Coeff. & ASPL & Diameter \\ 
\hline\hline
Western & $(497, 565)$ & $0.619$ & $3.27$ & $7$ \\ \hline
Eastern & $(575, 619)$ & $0.508$ & $2.91$ & $7$ \\ \hline
\end{tabular}
\end{table*}
%%%%%%%%%%%%%%%%%%%%%%%%%%%%%%%%%%%%%%%%%%%%%%
We now present the details of the interconnection structure of Western and Eastern Interconnects computed using Algorithm~\ref{algo.interconnect}. Key details are summrized in Table~\ref{interconnect}. We note that the interconnection graphs of the Polish and Texas Interconnect are small and unsuitable for analysis. The topological properties of the interconnection networks is substantially different from the properties of the decomposed networks presented in Section~\ref{sec:characterization}.
While the average clustering coefficient of these networks are higher, the average shortest path and diameter are significantly smaller (relative to decomposed networks). The fundamental differences in the topological structure is also evident from the visualization of the two systems presented in Figure~\ref{fig:interconnect}.

%%%%%%%%%%%%%%%%%%%%%%%%%%%%%%%%%%%%%%%%%%%%%%%%%%%%%%%%%%%%%%%%%%%%%%%%%%
%%%%%%%%%%%%%%%%%%%%%%%%%%%%%%%%%%%%%%%%%%%%%%%%%%%%%%%%%%%%%%%%%%%%%%%%%%

\begin{figure*}[h!]
%\begin{wrapfigure}{r}{0.5\textwidth}
  \begin{center}
   \includegraphics[width=0.50\textwidth]{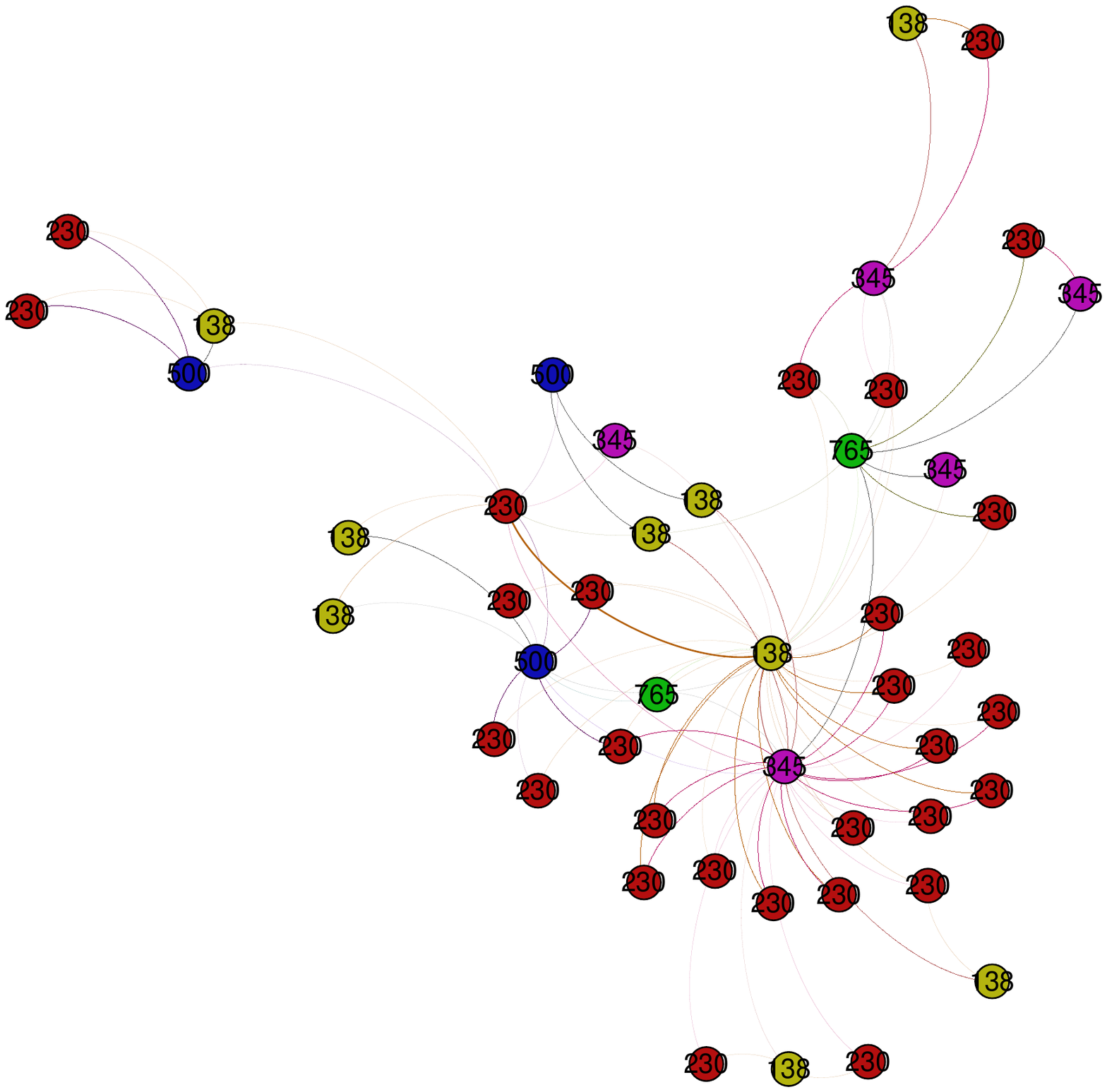}
   \end{center}
   \caption{Rendering of the 2-core of the Eastern interconnection network. Color code is as follows: $138$ -- yellow green, $230$ -- red, $345$ -- purple, $500$ -- blue, and $765$ -- green.}
  \label{fig:interconnect2core}
%\end{wrapfigure}  
\end{figure*}

From the interconnection graphs we observe a hierarchical nature of the manner in which different networks at different voltages interconnect with each other. The lowest voltage levels are used for distribution purposes and generally form degree-one vertices in the interconnection networks. The remaining vertices (networks of higher voltage) form a hierarchical structure spanning the network. In order to highlight this feature, we present a $2$-core of the interconnection network of the  Eastern Interconnect in Figure~\ref{fig:interconnect2core}. The $k$-core of a graph is a subgraph of the graph such that each vertex has degree $k$ or more. 
%%%%%%%%%%%%%%%%%%%%%%%%%%%%%%%%%%%%%%%%%%%%%%%%%%%%%%%%%%%%%%%%%%%%%%%%%%
%%%%%%%%%%%%%%%%%%%%%%%%%%%%%%%%%%%%%%%%%%%%%%%%%%%%%%%%%%%%%%%%%%%%%%%%%%

We will conclude this section by observing that the topological structure of the interconnection network is significantly different from the topological structure of the decomposed networks operating at the same voltage level. Further, there exists a hierarchical structure based on the voltage levels to the manner in which different vertices are linked to one another. 

\section{Random Graph Models}
\label{sec:RandomGraphModel}
An important research topic in the study of electric power grids is the ability to 
synthetically generate graphs that match the characteristics of real-world
power grids.  
We refer the reader to~\cite{Cotilla-Sanchez-Compare} for a detailed analysis of 
different random graph models and their comparison to real-world datasets. 
We summarize key findings from previous work in Section~\ref{sec:related}. 
In this section, we present two random graph models to generate 
graphs that match the characteristics of graphs at a given voltage level that
were presented in Section~\ref{sec:characterization}.
We then present graph models that match the characteristics of the interconnection
structure that we presented in Section~\ref{sec:interconnection}.
By combining the two separate models, we propose that synthetic graphs
can be generated to match the overall characteristics of real-world power grids.

We first present two random graph models to generate graphs at a given voltage. 
These two models can be seen as a variant of the random geometric graphs~\cite{penrose1}. 
A $d$-dimensional random geometric graph, $G(n, r)$, is a graph formed by 
randomly placing $n$ vertices in a $d$-dimensional space and adding an edge between
pairs of vertices whose Euclidean distance is less than or equal to $r$. 
Sparsity and structure of the generated graph can be controlled by varying the distance 
parameter $r$. We note here that we experimented with several random graph models and 
found that the random geometric graph models to be the most promising~\cite{Cotilla-Sanchez-Compare}. 
Here, we present two such models. The first model is given in 
Algorithm~\ref{algo.simpleMinDist}. 
The algorithm takes the size of the graph (number of vertices and edges) as the 
input, and generates a random graph as the output. The algorithm starts generating $n$ 
two-dimensional points uniformly at random (Line $3$ in Algorithm~\ref{algo.simpleMinDist}).
For each vertex $i$, a set of $\lfloor \frac{m}{n} \rfloor$ neighbors are chosen by 
minimizing the Euclidean distance as follows:
\begin{eqnarray}
\min_{j} &  & (x_{i}-x_{j})^{2}+(y_{i}-y_{j})^{2}\label{eq:mindist}\\
\mbox{s.t.} &  & j\notin adj(i)\nonumber, 
\end{eqnarray}
where $adj(i)$ represents the adjacency set of vertex $i$. 
An additional edge is generated with with probability 
$P=(\frac{m}{n} - \lfloor \frac{m}{n} \rfloor)$ (Line $6$ in Algorithm~\ref{algo.simpleMinDist}). 
 
%%%%%%%%%%%%%%%%%%%%%%%%%%%%%%%%%%%%%%%%%%%%%%%%%%%%%%%%%%%%%%%%%%%%%%%%%%%%
\begin{algorithm}[!htb]
\caption{Simple minimum-distance algorithm. {\bf Input:} Parameters for the number of 
vertices and edges ($n$ and $m$). {\bf Output:} A random graph $G$.}
\label{algo.simpleMinDist}
\begin{algorithmic}[1]
\Procedure{Simple-MinDist-Graph}{$(n,m)$, $G$}
\For {$i=1:n$} 
	\State Randomly generate planar coordinates for $i$ ($x_{i}$, $y_{i}$) from a uniform distribution
\EndFor
\For {$i=1:n$}
	\State Generate $\lfloor \frac{m}{n} \rfloor$ edges ($i,j$)
by iteratively selecting $j$ to minimize the Euclidean distance between
$i$ and $j$ using Equation~\ref{eq:mindist}

	\State Generate an additional edge with probability $P=(\frac{m}{n} - \lfloor \frac{m}{n} \rfloor)$ 
	using Equation~\ref{eq:mindist}	
\EndFor
\EndProcedure
\end{algorithmic}
\end{algorithm}
%%%%%%%%%%%%%%%%%%%%%%%%%%%%%%%%%%%%%%%%%%%%%%%%%%%%%%%%%%%%%%%%%%%%%%%%%%%%%%%%%%%%%

Graphs generated by Algorithm~\ref{algo.simpleMinDist} do not match 
all the desired properties of power grids. Therefore,  we consider
an adapted version of this algorithm that modifies how vertices get connected 
using the distance function given by Equation~\ref{eq:mindist}. 
The modification is driven by a bisection cost $c_{b}$. 
Unlike the previous algorithm, new nodes can either be connected to an existing node 
through a new edge, or it can be placed near an existing edge and become a bisecting 
node for that edge.
This change is motivated by economical considerations in the construction of 
power transmission networks.
For example, consider the construction of a new town that needs to
connect to the power grid. The town can either be connected via a
transmission line or feeder to an existing substation, or a new substation
could be built, bisecting an existing transmission line. 
With this modification, the min-dist selection criteria (Eq. \ref{eq:mindist})
is modified as follows:
\begin{eqnarray}
\min &  & (C_{1},C_{2})\nonumber \\
\mbox{s.t.} &  & C_{1}=\min_{j\notin adj(i)}(x_{i}-x_{j})^{2}+(y_{i}-y_{j})^{2}\label{eq:mindist2}\\
 &  & C_{2}=\min_{k\in\{1...m\}}d(i\rightarrow e_{k})+c_{j},\nonumber 
\end{eqnarray}
where $d(i\rightarrow e_{k})$ is the distance between point $i$ and the nearest point 
along line segment $e_{k}$, and $c_{j}$ is an exogenously selected bisection cost. 
If $C_{1}$ is smaller than $C_{2}$, then the algorithm creates a new edge 
$(i, j)$. Otherwise, the new node bisects the existing edge $e_{k}$.

\begin{figure*}[h!]
\subfigure[Interconnection graph for Western Interconnect]
{\includegraphics[width=0.45\textwidth]{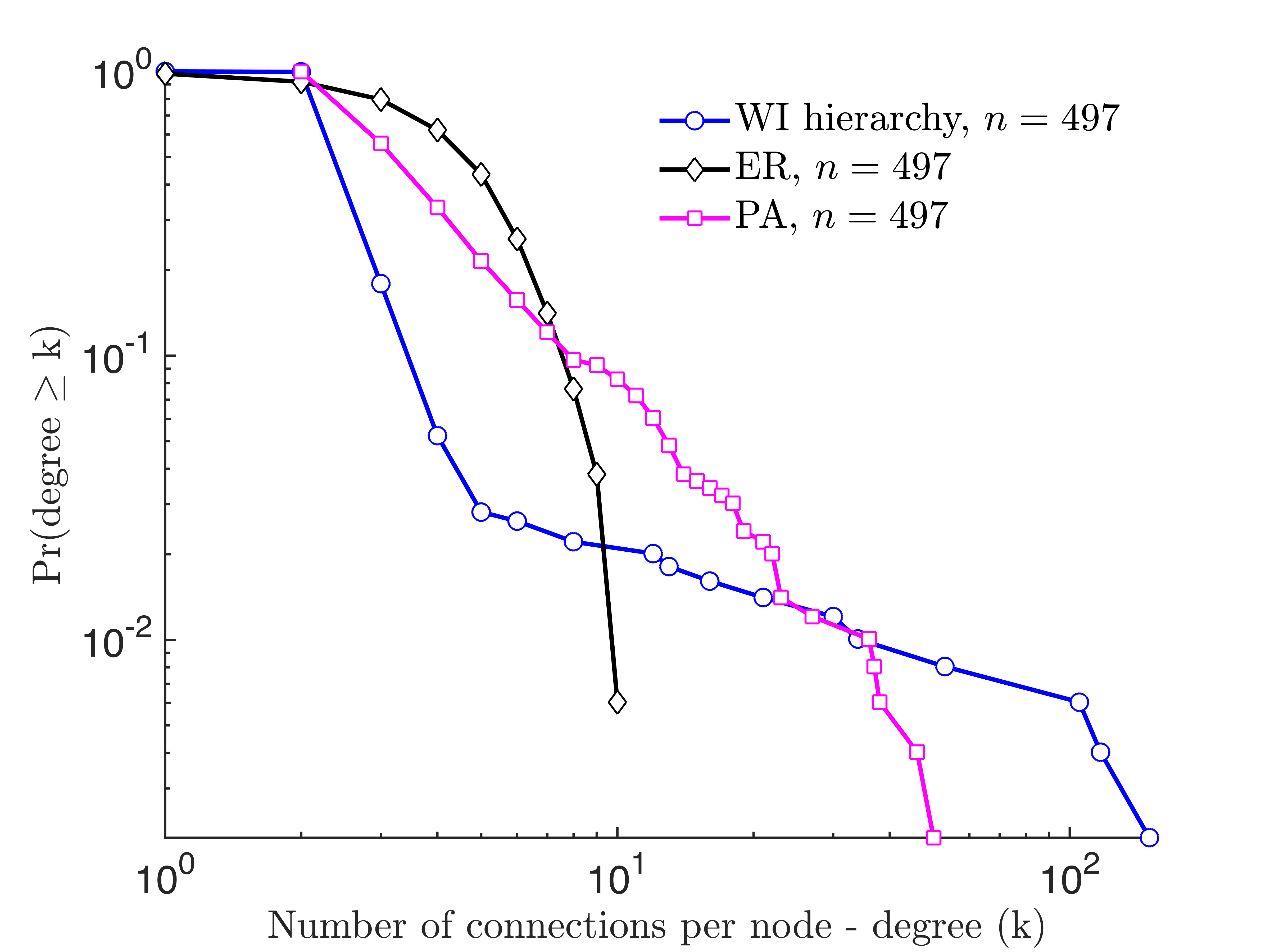}}
\subfigure[Interconnection graph for Eastern Interconnect]
{\includegraphics[width=0.45\textwidth]{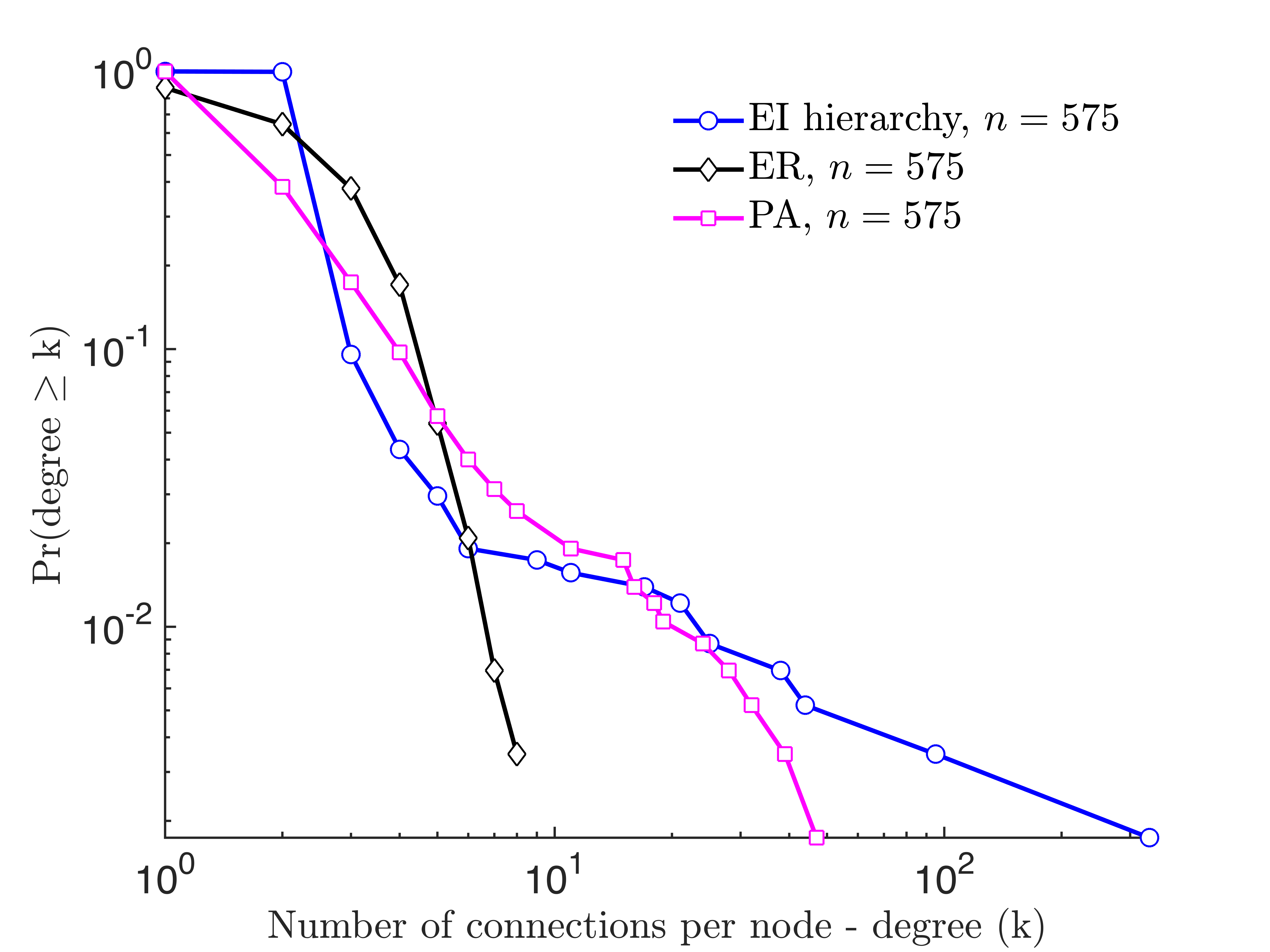}}
\caption{Cumulative degree distribution of the interconnection graphs of the Western and Eastern Interconnects,
along with degree distributions from random graphs -- Erd\H{o}s-R\'{e}nyi (ER) and preferential attachment (PA).}
\label{fig:degDistInt}
\end{figure*}

The second step in generating synthetic transmission networks is to generate random graphs
to match the characteristics of the interconnection graphs. We propose the use of 
preferential attachment (PA) model for this purpose. 
We generate the preferential attachment (PA) graphs using the algorithm proposed in 
~\cite{Barabasi}. Since we aim to generate graphs of a given size (number of vertices,
$n$, and edges $m$), we modify the algorithm in \cite{Barabasi} to allow for a 
fractional average degree $<k>$. 
For each new vertex $i$, $\lfloor \frac{m}{n} \rfloor$ edges are added by linking $i$ 
to an existing vertex $j$, where $j$ is chosen randomly from the probability distribution 
$P=k_j / \sum_{c=1}^{n}k_c$, where $k_c$ is the degree of vertex $c$. 
Further, an additional edge  $(i,j)$ is added with probability $\frac{m}{n} - 1$ to 
generate a preferential attachment graph with $n$ vertices and approximately $m$ edges.
We generate Erd\H{o}s-R\'{e}nyi (ER) random graphs by randomly adding edges between a 
pair of vertices until there are exactly $m$ edges in the graph.
Further details are provided in ~\cite{Cotilla-Sanchez-Compare}.
In Figure~\ref{fig:degDistInt}, we provide the cumulative degree distribution of the 
interconnection graphs of the Western and Eastern Interconnects (described in 
Section~\ref{sec:interconnection}) along with degree distributions from random graphs 
-- preferential attachment (PA) and Erd\H{o}s-R\'{e}nyi (ER) of the same sizes.
We observe that preferential attachment provides a better fit to the degree
distribution of interconnection graphs.
%%%%%%%%%%%%%%%%%%%%%%%%%%%%%%%%%%%%%%%%%%%%%%%%%%%%%%%%%%%%%%%%%%%%%%%%%%%%%%%%%%%%%
%%%%%%%%%%%%%%%%%%%%%%%%%%%%%%%%%%%%%%%%%%%%%%%%%%%%%%%%%%%%%%%%%%%%%%%%%%%%%%%%%%%%%

We conclude this section by noting that accurate models for synthetic recreation of transmission 
networks should include several aspects such as geography, cost of adding vertices and 
edges, size and distribution of power transformers, and the inherent hierarchy 
of networks of different voltage ratings. 
While we present a multi-step process involving different random graph models, 
careful construction and validation is part of our work in the near future.

\section{Related Work}
\label{sec:related}
Many researchers have explored the use of random graphs to model power grid
\cite{Cotilla-Sanchez-Compare,H2006,HBCB2010,Pagani,S2005,WST2009}, but in each case
the resulting random graph does not fully model the structure of interest. In
most cases, authors attempt to match graph characteristics from power grid
networks, such as degree distribution, average path length, and clustering
coefficient, in their random models.

In \cite{H2006} the author examines how vulnerability tests on the Western
Interconnect (WECC) and the Nordic grid, and then compare with the same tests on an
Erd\H{o}s-R\'enyi (ER) random graph and the scale-free model of Barab\'asi
and Albert \cite{BA1999}. They first note that the clustering coefficient and
average path length of the real graphs are significantly larger than that of
both types of random graphs. When comparing vulnerability to failure of
generators (i.e., removal of vertices) they see that the real networks show
more susceptibility than the random networks.

The authors of \cite{HBCB2010} also show that the topological properties of
the Eastern United States power grid and the IEEE 300 system differ
significantly from random networks created using preferential-attachment,
small-world, and ER models. They provide a new model, the minimum distance
graph, which more closely matches measured topological characteristics like
degree distribution, clustering coefficient, diameter, and assortativity of
the real networks. We build on these ideas in our work. Extensions of the
minimum-distance model are discussed in Section~\ref{sec:RandomGraphModel}.

In a more recent paper by some of the same authors \cite{HBCB2010}
they look more specifically at both the electrical and topological
connectivity of three North American power infrastructures. They again
compare these with random, preferential-attachment, and small-world networks
and see that these random networks differ greatly from the real power
networks. They propose to represent electrical connectivity of power systems
using electrical distances rather than physical connectivity and geographic
connections. In particular, they propose a distance based on sensitivity
between active power transfers and nodal phase angle differences. Electrical
distance is calculated as a positive value for all pairs of vertices which
then yields a complete weighted graph. Therefore, to make it more comparable
to a geographic network the authors use a threshold value and keep those
edges which are below the threshold.

Pagani and Aiello, in \cite{Pagani}, present a survey of the most relevant
work using complex network analysis to study the power grid. In this survey
they remark that most of the networks studied were High Voltage from either
North America, Europe, or China. They remark that most of the surveyed work show
that degree distribution is exponential. The geography of the country seems
to be important as the results differ somewhat between countries with
differing geographies. One point of agreement between all studies is that the
power grid is resilient to random failures but extremely vulnerable to
attacks targeting nodes with high degree or high betweenness scores.

Finally, we cite the work in \cite{WST2009} in which the authors characterize
many graph measures in the context of power grid graphs. As an example, they show
that power grids are sparsely connected, the degree distribution has an
exponential tail, and the line impedance has a heavy-tailed distribution.
Based on their findings they propose an algorithm to generate random power
grids which features the same topology and electrical characteristics
discovered from real data. They, like us, take a hierarchical approach to
generating synthetic power networks. However, their approach differs by
looking at geographic zones, whereas our approach is to break up the network
by voltage level. Our approach requires no geographic knowledge of the
system and leads to a systematic approach for annotating the nodes and edges
with different electrical properties such as voltage ratings.

\section{Conclusions and Future Work}
\label{sec:conclusion}
Graph-theoretic analysis of power grids has often resulted
in misleading conclusions. In this paper, we hypothesized that
one of the reasons for such conclusions is the inability to 
account for heterogeneity in a power grid. 
We therefore developed a new method for decomposition 
based on nominal voltage rating, and using power 
transformers to generate a network-of-networks model of
power transmission networks.  
While the individual networks operating at the same voltage
level are characterized by exponential degree distribution,
large values for average shortest-path length and diameter, and
low clustering coefficient; the interconnection structure 
representing the interconnection of these networks is 
characterized by non-exponential degree distribution,
small values for average shortest-path length and diameter, 
and relatively higher clustering coefficient.

Consequently, we proposed two random graph methods for 
synthetic generation of power transmission networks. Our 
approach not only models the topology, but also provides a
method for annotating the network with voltage ratings so 
that the electrical properties of a grid can be modeled 
correctly. To the best of our knowledge, this approach is novel.

The resulting characterization of power transmission networks
and the ability to synthetically recreate networks has 
important implications for studying the vulnerability of 
power systems, evolution of power networks, and transmission 
expansion planning.
Thus, the ideas proposed in this paper hold the potential to 
significantly improve our understanding of several aspects of 
electric infrastructure networks, as well as other critical 
infrastructure networks.

\section*{Acknowledgments}
This work was supported in part by the Applied Mathematics Program of the Office
of Advance Scientific Computing Research within the Office of Science of
the U.S. Department of Energy (DOE). Pacific Northwest National Laboratory
(PNNL) is operated by Battelle for the DOE under Contract
DE-AC05-76RL01830.


\begin{thebibliography}{10}

\bibitem{hines-chaos2010}
P.~Hines, E.~Cotilla-Sanchez, and S.~Blumsack, ``{Do topological models provide
  good information about electricity infrastructure vulnerability?},'' {\em
  Chaos: An Interdisciplinary Journal of Nonlinear Science}, vol.~20, no.~3,
  p.~033122, 2010.

\bibitem{DeMarco}
D.~Cheverez-Gonzalez and C.~DeMarco, ``Admissible locational marginal prices
  via laplacian structure in network constraints,'' {\em Power Systems, IEEE
  Transactions on}, vol.~24, pp.~125--133, Feb 2009.

\bibitem{Xu-2010}
G.~Xu, {\em Controlled Islanding Algorithms and Demonstrations on the Wecc
  System}.
\newblock PhD thesis, Tempe, AZ, USA, 2010.

\bibitem{Anderson}
J.~Anderson and A.~Chakrabortty, ``Graph-theoretic algorithms for pmu placement
  in power systems under measurement observability constraints,'' in {\em Smart
  Grid Communications (SmartGridComm), 2012 IEEE Third International Conference
  on}, pp.~617--622, Nov 2012.

\bibitem{Cotilla-Sanchez-Compare}
E.~Cotilla-Sanchez, P.~Hines, C.~Barrows, and S.~Blumsack, ``Comparing the
  topological and electrical structure of the north american electric power
  infrastructure,'' {\em Systems Journal, IEEE}, vol.~6, pp.~616--626, Dec
  2012.

\bibitem{Zimmerman:2011}
R.~D. Zimmerman, C.~E. Murillo-S{\'a}nchez, and R.~J. Thomas, ``Matpower:
  Steady-state operations, planning and analysis tools for power systems
  research and education,'' {\em IEEE Transactions on Power Systems}, vol.~26,
  pp.~12--19, Feb. 2011.

\bibitem{penrose1}
M.~Penrose, {\em Random Geometric Graphs}.
\newblock Oxford University Press, 2003.

\bibitem{Barabasi}
A.-L. Barab\'{a}si and R.~Albert, ``Emergence of scaling in random networks,''
  {\em Science}, vol.~286, no.~5439, pp.~509--512, 1999.

\bibitem{H2006}
A.~J. Holmgren, ``Using graph models to analyze the vulnerability of electric
  power networks,'' {\em Risk Analysis}, vol.~26, no.~4, pp.~955--969, 2006.

\bibitem{HBCB2010}
P.~Hines, S.~Blumsack, E.~C. Sanchez, and C.~Barrows, ``The topological and
  electrical structure of power grids,'' in {\em Proceedings of the 43rd
  {H}awaii International Conference on System Sciences}, 2010.

\bibitem{Pagani}
G.~A. Pagani and M.~Aiello, ``The power grid as a complex network: A survey,''
  {\em Physica A: Statistical Mechanics and its Applications}, vol.~392,
  no.~11, pp.~2688--2700, 2013.

\bibitem{S2005}
K.~Sun, ``Complex networks: A new method of research in power grid,'' in {\em
  Proceedings of the 2005 IEEE/PES Transmission and Distribtution Conference \&
  Exhibition}, 2005.

\bibitem{WST2009}
Z.~Wang, A.~Scaglione, and R.~J. Thomas, ``On modeling random topology power
  grids for testing decentralized network control strategies,'' in {\em
  Proceedings of the 1st IFAC Workshop on Estimation and Control of Networked
  Systems}, pp.~114--119, 2009.

\bibitem{BA1999}
A.~L. Barab\'asi and R.~Albert, ``Emergence of scaling in random networks,''
  {\em Science}, vol.~286, pp.~509--512, 1999.

\end{thebibliography}
\end{document}